\documentclass[10pt, journal,twoside]{IEEEtran} 

\ifCLASSOPTIONcompsoc
\usepackage[nocompress]{cite}
\else
\usepackage{cite}
\fi

\usepackage{epsfig}
\usepackage{amsmath}
\usepackage{amssymb}
\usepackage{cases}
\usepackage{hhline}
\usepackage{times} 
\usepackage[font=footnotesize,labelfont=bf]{caption}
\usepackage{caption}
\usepackage{algorithm}
\usepackage{algorithmic}
\usepackage[T1]{fontenc}
\usepackage{color}
\usepackage{tabularx}
\usepackage{booktabs}
\usepackage{makecell}
\usepackage{multirow}
\usepackage{float}
\usepackage{subfig}
\usepackage{graphicx}
\usepackage{url}

\begin{document}
\title{Label Leakage Attacks in Machine Unlearning: A Parameter and Inversion-Based Approach}

\author{Weidong Zheng, Kongyang Chen, Yao Huang, Yuanwei Guo, Yatie Xiao
\IEEEcompsocitemizethanks{
\IEEEcompsocthanksitem{Weidong Zheng and Yatie Xiao are with School of Computer Science and Cyber Engineering, Guangzhou University, Guangzhou 510335, China.} 
\IEEEcompsocthanksitem{Kongyang Chen and Yao Huang are with School of Artificial Intelligence, Guangzhou University, Guangzhou 510006, China. Kongyang Chen is also with Pazhou Lab, Guangzhou 510006, China. E-mail: kychen@gzhu.edu.cn.}
\IEEEcompsocthanksitem{Yuanwei Guo is with Guangzhou Institute of Internet of Things, Guangzhou 511462, China.} 
}}

\IEEEtitleabstractindextext{
\begin{abstract}
With the widespread application of artificial intelligence technologies in face recognition and other fields, data privacy security issues have received extensive attention, especially the \textit{right to be forgotten} emphasized by numerous privacy protection laws. Existing technologies have proposed various unlearning methods, but they may inadvertently leak the categories of unlearned data. This paper focuses on the category unlearning scenario, analyzes the potential problems of category leakage of unlearned data in multiple scenarios, and proposes four attack methods from the perspectives of model parameters and model inversion based on attackers with different knowledge backgrounds. At the level of model parameters, we construct discriminative features by computing either dot products or vector differences between the parameters of the target model and those of auxiliary models trained on subsets of retained data and unrelated data, respectively. These features are then processed via k-means clustering, Youden’s Index, and decision tree algorithms to achieve accurate identification of the forgotten class. In the model inversion domain, we design a gradient optimization-based white-box attack and a genetic algorithm-based black-box attack to reconstruct class-prototypical samples. The prediction profiles of these synthesized samples are subsequently analyzed using a threshold criterion and an information entropy criterion to infer the forgotten class. We evaluate the proposed attacks on four standard datasets against five state-of-the-art unlearning algorithms, providing a detailed analysis of the strengths and limitations of each method. Experimental results demonstrate that our approach can effectively infer the classes forgotten by the target model.
\end{abstract}
}

\maketitle
\IEEEdisplaynontitleabstractindextext
\IEEEpeerreviewmaketitle

\section{Introduction}\label{sec:introduction}

Driven by deep learning, artificial intelligence technologies have been widely applied in various fields such as autonomous driving, financial risk assessment, medical disease diagnosis, and biometric recognition. Their superior performance heavily relies on in-depth mining of massive training data. However, with the widespread deployment of data-driven models, the risk of sensitive privacy leakage in training data has become increasingly prominent. Existing studies have shown that attack methods such as membership inference attacks can steal private data from trained models\cite{shokri2017membership}. Therefore, exploring potential data privacy issues has become a common focus of both academia and industry. Meanwhile, numerous privacy protection laws emphasize the \textit{right to be forgotten}, which entitles users to proper control over their personal data and explicitly requires service providers to completely remove personal data from artificial intelligence models under specific circumstances. As an important enabler for implementing the \textit{right to be forgotten}, machine unlearning techniques aim to efficiently delete specified data from artificial intelligence models.

In recent years, extensive research has been conducted on machine unlearning, spanning from traditional machine learning models like logistic regression to sophisticated deep neural networks(DNNs) such as ResNet. Regarding technical approaches, two main paradigms have emerged: system-level methods, including SISA\cite{bourtoule2021machine} and ARCANE\cite{yan2022arcane}, and algorithm-level techniques, such as those based on Fisher information\cite{golatkar2020eternal, golatkar2020forgetting, golatkar2021mixed} and influence functions\cite{guo2019certified, koh2017understanding, dwork2006our, wu2022puma}. Collectively, these advances provide essential technical support for enforcing personal privacy rights, notably the "right to be forgotten."

Although machine unlearning was originally proposed to preserve data privacy, it does not guarantee that data privacy remains secure after unlearning\cite{ye2025enhancing}. Chen et al. \cite{chen2021machine} investigated membership inference attacks against machine unlearning, which determine whether a data sample is a member of the target model's training set by exploiting the prediction differences of the model on the forgotten data before and after unlearning. Compared with classical membership inference attacks, this method achieves promising performance on well-generalized models. Hu et al.  \cite{hu2024learn} were the first to propose the use of an unlearning inversion attack to infer the label of the unlearned data. However, this method requires simultaneous access to both the pre-unlearning and post-unlearning models. Such an assumption is often unrealistic in practical settings, as the outdated model is typically deleted or discarded once the unlearning procedure is completed.At present, there are few studies on privacy leakage against machine unlearning, and more work is needed to explore the mechanisms of such privacy leakage to better safeguard data privacy.

\begin{table}[htbp]
  \centering
  \caption{Attack Methods Corresponding to Each Scenario in This Paper}
  \label{tab:attack_methods}
  \footnotesize
\setlength{\tabcolsep}{1pt}
 \renewcommand{\arraystretch}{1}
  \begin{tabular}{l l c c c}
    \toprule
    \textbf{Method} &\textbf{Classification} & \makecell{ \bfseries White-\\ \bfseries box} & \makecell{\bfseries Black-\\ \bfseries box} & 
    \makecell{\bfseries Access to Small\\ \bfseries Training Set} \\
    \midrule
    \multirow{2}{*}{\makecell[l]{Model Parameter \\ Dot Product}} 
      & k-means & \checkmark & & \checkmark \\
    \cmidrule(lr){2-5}
      & Youden’s Index & \checkmark & & \checkmark \\
    \midrule
    \makecell[l]{Model Parameter \\ Difference}
      & Decision Tree & \checkmark & & \checkmark \\
    \midrule
    \multirow{2}{*}{\makecell[l]{Model Inversion via \\ Gradient Optimization}} 
      & Threshold & \checkmark & & \\
    \cmidrule(lr){2-5}
      & Entropy & \checkmark & & \\
    \midrule
    \multirow{2}{*}{\makecell[l]{Model Inversion via \\ Genetic Algorithm}} 
      & Threshold & & \checkmark & \\
    \cmidrule(lr){2-5}
      & Entropy & & \checkmark & \\
    \bottomrule
  \end{tabular}
\end{table}

This paper focuses on the class-level unlearning scenario, analyzes the potential privacy leakage issues inherent in model unlearning under this scenario, and for the first time investigates the label leakage problem of the data to be forgotten during unlearning from the perspectives of both model parameters and model inversion, using only the unlearned model.As shown in Table~\ref{tab:attack_methods}, given that model parameters reflect the statistical distribution and sample information of the training set, we achieve effective inference of the target forgotten class by comparing the parameter differences among models trained on a small training data, models trained on irrelevant data, and the target unlearned model. Corresponding classification strategies are designed based on this analysis. Concurrently, we provide rigorous mathematical derivations to prove the validity of the approach, establishing a solid theoretical foundation. Furthermore, by conducting model inversion attacks on the target model, we observe a distinct difference in the predicted probability distributions between the inverted data of the forgotten class and those of other classes. Building upon this observation, we design class inference attacks tailored for both white-box and black-box scenarios.

The main contributions of this paper are summarized as follows:

\begin{enumerate}
    \item We conduct a systematic study on the core issue of label leakage of data to be forgotten during model forgetting across multiple scenarios. Specifically, we propose a white-box attack method based on the dot product and difference of model parameters. For scenarios where the attacker possesses a small amount of training data, label screening methods, including k-means clustering, Youden’s Index, and decision trees, are designed accordingly. 
    \item We introduce feature inversion methods based on gradient optimization and genetic algorithms, respectively, for both white-box and black-box scenarios in which the attacker cannot access any training data, to obtain the model's prediction probabilities for each class. In addition, threshold-based and entropy-based label screening criteria are proposed, enabling precise identification of labels corresponding to the data to be forgotten.
    \item We evaluate our attacks on five classical model forgetting algorithms using standard datasets including MNIST, Fashion-MNIST, SVHN, and CIFAR-10, achieving high ASR while also analyzing the advantages and limitations of each method.
\end{enumerate}

\section{Related Work}\label{sec:Related Work}

With the rapid evolution of information technology and the widespread adoption of intelligent applications, data has become a critical factor driving societal progress, yet it simultaneously triggers severe crises in data ethics and privacy security. In recent years, tech giants such as Apple, Google, and Microsoft have frequently been embroiled in legal disputes concerning data abuse and privacy violations, highlighting the urgent need to protect individual rights in the digital economy era. To address this challenge, major global economies have established comprehensive data compliance governance frameworks. Representative legislation includes the European Union's General Data Protection Regulation (GDPR) \cite{GDPR2016}, the California Consumer Privacy Act (CCPA) in the United States \cite{CCPA2018}, and Data Security Law of the People's Republic of China \cite{DataSecurityLaw2021}. Notably, the GDPR is the first to legally establish the \textit{right to be forgotten}, granting data subjects the statutory right to request the erasure of their personal data by data controllers. It is important to note that in the current landscape where artificial intelligence and machine learning technologies are deeply pervasive, the scope of this right has extended beyond the traditional deletion of database records to encompass algorithm models trained on personal data and their associated application scenarios. This evolution presents novel technical challenges and legal interpretation dilemmas for existing data governance paradigms.

Following the legal implementation of the \textit{right to be forgotten} in data-driven systems, the need to precisely remove the influence of specific data from complex machine learning models has catalyzed the emergence of \textit{machine unlearning} as a new research direction \cite{cao2015towards}. Unlike conventional database deletion operations, machine unlearning requires the erasure of feature representations of designated samples within the parameter space, restoring the model's output distribution to a state indistinguishable from one where the data was never learned. Although performing full retraining on the remaining data serves as the baseline method for achieving this objective, its prohibitively high computational cost severely limits its applicability in dynamic data environments. Consequently, the academic research focus has gradually shifted towards incremental, perturbative, or approximate unlearning strategies, aiming to approximate the effectiveness of full retraining at a lower computational cost.

\textbf{Machine Unlearning Algorithms:} Machine Unlearning aims to eliminate the residual influence of specific training data on model parameters and outputs, forcing the model's behavior to converge to a state of \textit{never having learned that data}. These algorithms are broadly categorized into Exact Unlearning and Approximate Unlearning. Exact Unlearning refers to the complete removal of specified data information from the model \cite{bourtoule2021machine, yan2022arcane}. In contrast, Approximate Unlearning relaxes the strict requirement for result consistency, pursuing statistical indistinguishability where the post-unlearning model shows no significant distributional deviation from a retrained model. Recent studies have proposed various techniques to achieve approximate unlearning, including Influence Functions for quantifying parameter impact \cite{guo2019certified, koh2017understanding, dwork2006our, wu2022puma}, Fisher Information Matrix-based perturbations \cite{golatkar2020eternal, golatkar2020forgetting, golatkar2021mixed}, training trajectory replay \cite{graves2021amnesiac, wu2020deltagrad}, model sparsification followed by pruning \cite{jia2023model}, Hessian-guided overfitting approximation, and label noise injection \cite{zheng2026accurate}. The application scenarios for these techniques have expanded to Graph Neural Networks \cite{ye2025towards}, Federated Learning \cite{pan2025feature, wang2026blindu}, and Large Language Models  \cite{wang2025selective}.

\textbf{Privacy and Security in Machine Unlearning:} Regarding security issues in the unlearning process, Chen et al. \cite{chen2021machine} first demonstrated that the unlearning process itself could become a new source of privacy leakage. By comparing the behavioral differences of models before and after the unlearning operation, they successfully identified the membership attributes of data slated for forgetting using Membership Inference Attacks, challenging the traditional assumption that \textit{unlearning equals privacy protection}. To further investigate the fragility of unlearning mechanisms, Marchant et al. \cite{marchant2022hard} designed data poisoning attacks against linear models from a cost-sensitive perspective, proving that subtle feature tampering of samples to be forgotten could lead to a sharp escalation in model unlearning costs. Subsequently, Di et al. \cite{di2022hidden} proposed Camouflage Poisoning Attacks, indicating that attackers could insert both poisoned data and mitigating data into the training set, activating malicious effects during the unlearning phase by removing the mitigating data. This series of studies demonstrates that machine unlearning exhibits significant vulnerability to malicious attacks.

Research on the security analysis and vulnerability of machine unlearning mechanisms includes Ye et al., who systematically explored Malicious Unlearning Attacks (MUEM) and Privacy Inference Attacks (MIEM) in ensemble models for the first time, revealing security risks where machine unlearning techniques in ensemble frameworks could be abused for covert poisoning or leaking labels of unlearned data \cite{ye2025enhancing}. Concurrently, Barez et al. pointed out fundamental limitations in current machine unlearning techniques regarding the control of AI capabilities, specifically noting that in dual-use knowledge management scenarios, unlearning may fail to completely eliminate a model's harmful capabilities \cite{barez2025open}. George et al. found severe instability in unlearning methods for text-to-image diffusion models, where fine-tuning a forgotten model causes the resurrection of forgotten concepts, indicating a lack of robustness in current techniques \cite{george2025illusion}. Hu et al. proposed unlearning inversion attacks, which first revealed that attackers could infer the features and labels of unlearned data by merely accessing the original and post-unlearning models, exposing privacy leakage vulnerabilities in the unlearning process \cite{hu2024learn}.

Although research on machine unlearning and its security has made progress, the issue of \textit{label leakage during model unlearning} remains underexplored. Therefore, focusing on DNNs, this paper systematically investigates the label leakage risks in both white-box and black-box scenarios under different unlearning methods.

\section{Problem Description}\label{Problem Description}

\subsection{Overview of the Attack Methodology}

This section first analyzes the privacy leakage issues inherent in model unlearning, followed by separate introductions to the threat model, attack targets, and attack objectives.

\subsection{Privacy Leakage Analysis of Model Unlearning}
The shift in the prediction probability distribution of a model concerning target data before and after unlearning harbors severe privacy leakage risks. Chen et al. \cite{chen2021machine} successfully conducted Membership Inference Attacks by comparing the model's output differences for data before and after the unlearning operation. However, this method critically relies on prior knowledge of the original training data distribution and white-box access to the pre-unlearning model parameters, conditions that are often infeasible in practical deployment scenarios. Given that model update mechanisms typically adhere to an overwrite deployment principle, where the unlearned model directly replaces the original, attackers can rarely obtain dual model snapshots before and after the state transition. Furthermore, while attackers might possess a small number of  training samples in some edge cases, the data targeted for forgetting is usually highly confidential and inaccessible externally. Therefore, launching data exfiltration attacks under the restricted condition of only being able to access the post-unlearning model constitutes a more realistic and challenging research problem. This paper is grounded precisely in such stringent scenarios, delving into the mechanisms of label information privacy leakage during class-level unlearning processes.

In the threat model of Class-level Unlearning, if an attacker possesses complete knowledge of the training data distribution, they can easily identify the forgotten target class by systematically probing the unlearned model. Specifically, the attacker can feed validation samples covering all classes into the target model and discriminate based on the significant performance disparity in predictions between the forgotten and retained classes. Although attackers might have participated in the data construction process or hold a few training samples in certain specific scenarios, in practical deployments, training data typically constitutes highly sensitive commercial secrets, making it difficult for attackers to acquire large-scale genuine training sets. A feasible strategy involves using this data to train a new model, exploiting the differences in model parameters between models trained on the full training data, the unlearned data, and the target model, thereby inferring the forgotten class. In a more stringent \textit{zero-shot} attack scenario, where the attacker has no access whatsoever to the model's original training data or even its internal architecture, Model Inversion Attacks  \cite{fredrikson2015model} provide a highly threatening solution. This method works by reconstructing representative samples in the input space in reverse from the target model's outputs and analyzing the feature distribution and semantic consistency of the generated samples, which can effectively determine whether a specific class has been deliberately removed during the unlearning process. The analysis above indicates that even under the extreme condition of inaccessible training data, class-level unlearning operations still face severe privacy leakage risks.

\subsection{Threat Model}
While existing studies exploit the state difference between the target model and the unlearned model for privacy attacks \cite{chen2021machine}, in practical application scenarios, compliance considerations typically mandate that the unlearning operation follows an instant overwrite principle. This means the original model is taken offline immediately after the update, fundamentally eliminating the attacker's possibility of obtaining dual-model differences. Given this, this paper focuses on the restricted scenario where only the single, post-unlearning model is accessible, to investigate the privacy leakage of forgotten data. The attacker's capabilities are assumed as follows:

\begin{enumerate}
    \item \textbf{White-box Attack (With Training Data)}: The attacker possesses full white-box access to the target model, including its architecture and all internal parameters. Additionally, the attacker holds a small number of  training samples that follow the same distribution as the training data.
    \item \textbf{White-box Attack (Without Training Data)}: The attacker also has complete visibility into the model's architecture and parameters but possesses no original training data. Attacks rely solely on public knowledge or synthetic data.
    \item \textbf{Black-box Attack}: The attacker has no knowledge of the model's internal structure and can only interact with the model via its query interface, obtaining the model's output, i.e., the prediction probability vector.
\end{enumerate}

Beyond the access-related assumptions above, this paper assumes the attacker possesses certain domain-specific public knowledge, such as prior information regarding the Cross-Entropy Loss function commonly used in classification tasks.

\subsection{Attack Target}

The target of the attacks in this paper is the deep neural network model produced in a Class-level Unlearning scenario. The attacker aims to infer the specific class information that has been removed from the model by the unlearning algorithm, through the analysis of model parameters and inversion techniques.

\subsection{Attack Objective}

The core research scenario of this paper is positioned within Class-level Unlearning, aiming to precisely identify the target class labels that have been removed from the model through inverse analysis. The harmfulness of such attacks manifests as a dual threat to both commercial entities and individual users: First, on the enterprise side, attackers can deduce internal operational dynamics by analyzing the traces left by model unlearning, leading to the leakage of trade secrets and significantly weakening the enterprise's core competitiveness and market advantage. Second, on the user side, if a user has already submitted a data deletion request, attackers can exploit the metadata traces left by class unlearning to establish a strong association between the user's identity and the forgotten data, thereby substantially infringing upon user privacy and nullifying the legal efficacy of the right to be forgotten.

\section{Our methods}\label{sec:Our methods}

\subsection{Label Inference Attack based on Model Parameters}
During the model learning process, a model continuously adjusts its parameters through the training set to learn the data features. Ultimately, the obtained model parameters inherently reflect the characteristics of the training set under a specific model architecture. Therefore, we attempt to identify whether certain specific data have been forgotten from the perspective of model parameters. To this end, we provide rigorous theoretical justification for this concept through mathematical principles and propose two attack methods based on model parameter similarity and difference.
\begin{figure*}[t]
    \centering
    \includegraphics[width=0.6\textwidth]{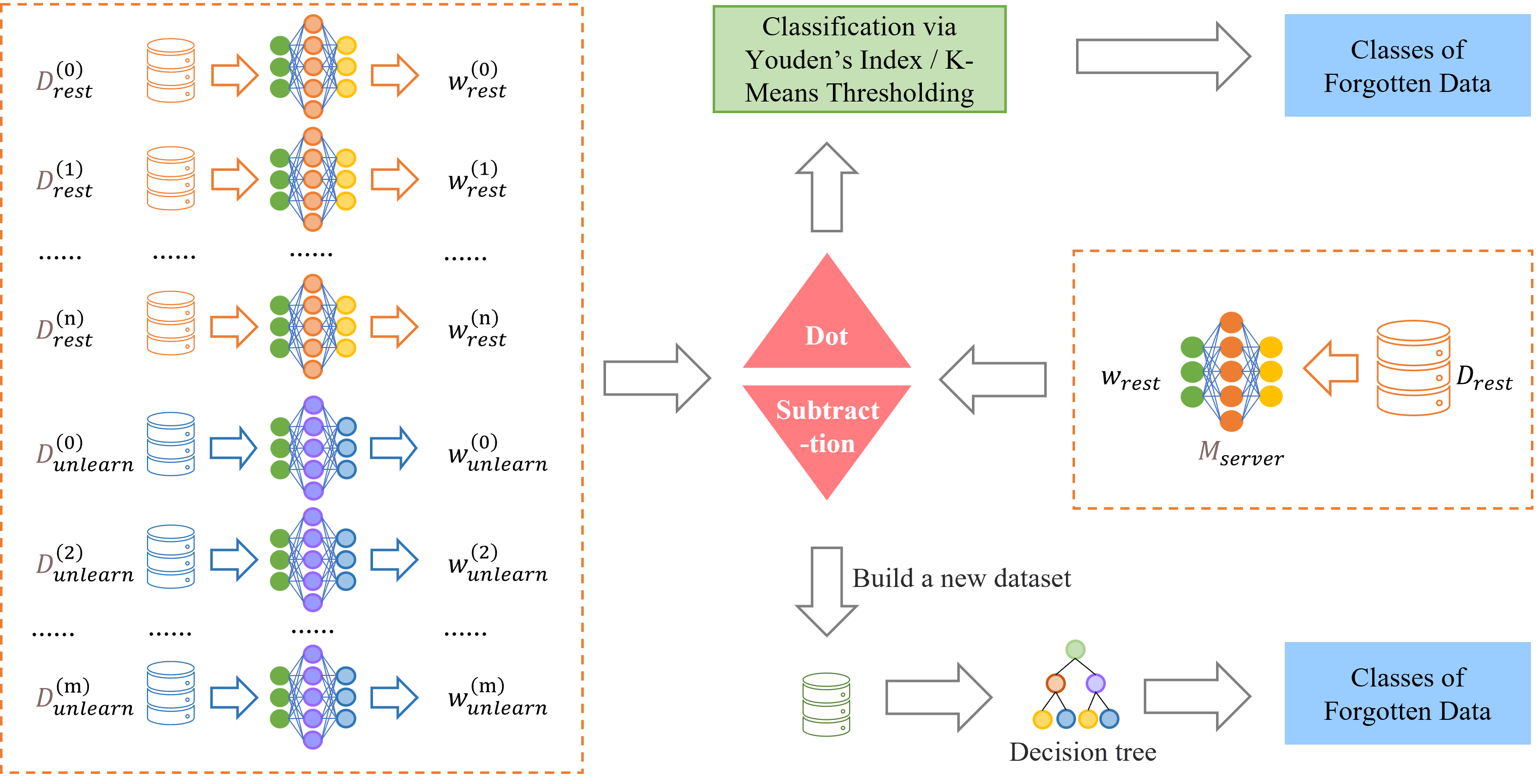} 
    \caption{Framework of the Label Inference Attack Based on Model Parameters}
    \label{fig:param_attack_flowchart}
\end{figure*}
\subsubsection{Attack Method based on Model Parameter Similarity}

For a dataset $D$, the original server model $M$ is obtained by training on $D$. Let $D_{\text{rest}} \subset D$, and $D_{\text{unlearn}} = D - D_{\text{rest}}$, where $D_{\text{unlearn}}$ is the data to be forgotten. While various model unlearning methods can be applied, the ideal outcome is to retrain a new model with $D_{\text{rest}}$ using the same architecture as $M$, resulting in $M_{\text{rest}}$, which is the model on the server after unlearning. We denote the parameters of its fully connected layer as a one-dimensional vector $w_{\text{rest}}$. Our objective is to infer the classes of the forgotten data $D_{\text{unlearn}}$ from $M_{\text{rest}}$, where $D_{\text{rest}}$ does not contain that classes. Note that $D_{\text{rest}}$ consists of many classes, i.e., $D_{\text{rest}} = D_{\text{rest}}^{(1)} + D_{\text{rest}}^{(2)} + \cdots + D_{\text{rest}}^{(n)}$. For simplicity in the following derivation, we denote one of these subsets as $D_{\text{rest}}'$.

We train multiple models, $M_{\text{rest}}^{(1)}, M_{\text{rest}}^{(2)}, \cdots, M_{\text{rest}}^{(n)}$, on $D_{\text{rest}}^{(1)}, D_{\text{rest}}^{(2)}, \cdots, D_{\text{rest}}^{(n)}$ respectively, using a model architecture identical to $M_{\text{rest}}$. Crucially, we freeze the parameters of the feature extraction layers, initializing them with those from $M_{\text{rest}}$, and only train the final classification layers. We denote one of these trained models as $M_{\text{rest}}'$, with its fully connected layer parameters as $w_{\text{rest}}'$.

Similarly, $D_{\text{unlearn}} = D_{\text{unlearn}}^{(1)} + D_{\text{unlearn}}^{(2)} + \cdots + D_{\text{unlearn}}^{(m)}$. We denote one subset as $D_{\text{unlearn}}'$. Employing the same method, we train models $M_{\text{unlearn}}^{(1)}, M_{\text{unlearn}}^{(2)}, \cdots, M_{\text{unlearn}}^{(m)}$ on $D_{\text{unlearn}}^{(1)}, D_{\text{unlearn}}^{(2)}, \cdots, D_{\text{unlearn}}^{(m)}$, respectively. The architecture and feature extraction initialization strategy remain identical. We denote one such model as $M_{\text{unlearn}}'$, with its fully connected layer parameters as $w_{\text{unlearn}}'$.

\subsection{Theoretical Analysis and Attack Formulation}

Considering the characteristics that$D_{\text{rest}}' \subset D_{\text{rest}}$ and $D_{\text{unlearn}}' \cap D_{\text{rest}} = \emptyset$, along with the decisive role of the dataset in determining the final model parameters, we attempt to launch an inference attack from the perspective of the fully connected layer parameters.

Let the mapping of the frozen feature extraction layers be $\phi: \mathcal{X} \rightarrow \mathbb{R}^d$, which outputs a $d$-dimensional vector $\phi(\mathbf{x})$ for an input data sample $\mathbf{x}$. Let the distributions of $D_{\text{rest}}$ and $D_{\text{unlearn}}'$ be $P_{\text{rest}}$ and $P_{\text{unlearn}}'$, respectively. Since the data volume of $D_{\text{rest}}'$ is small, it shares the same distribution as $D_{\text{rest}}$ and can be approximated as a sample drawn from $D_{\text{rest}}$.

Define the following expectations:
\begin{align}
\mathbb{E}_{\mathbf{x} \sim P_{\text{rest}} | \mathbf{x} \in D_{\text{rest}}} [\phi(\mathbf{x})] &= \mu_{\text{rest}}, \\
\mathbb{E}_{\mathbf{x} \sim P_{\text{rest}} | \mathbf{x} \in D_{\text{rest}}'} [\phi(\mathbf{x})] &= \mu_{\text{rest}}', \\
\mathbb{E}_{\mathbf{x} \sim P_{\text{unlearn}}' | \mathbf{x} \in D_{\text{unlearn}}'} [\phi(\mathbf{x})] &= \mu_{\text{unlearn}}'; \\
\mathbb{E}_{\mathbf{x} \sim P_{\text{rest}}' | \mathbf{x} \in D_{\text{rest}}'} [\phi(\mathbf{x}) \phi(\mathbf{x})^\mathsf{T}] &= \Sigma_{\text{rest}}', \\
\mathbb{E}_{\mathbf{x} \sim P_{\text{unlearn}}' | \mathbf{x} \in D_{\text{unlearn}}'} [\phi(\mathbf{x}) \phi(\mathbf{x})^\mathsf{T}] &= \Sigma_{\text{unlearn}}', \\
\mathbb{E}_{\mathbf{x} \sim P_{\text{rest}} | \mathbf{x} \in D_{\text{rest}}} [\phi(\mathbf{x}) \phi(\mathbf{x})^\mathsf{T}] &= \Sigma_{\text{rest}}.
\end{align}

We assume the Hessian matrix, the second-order derivative of the loss function, is positive definite. Let the expected risk be $\mathcal{L}(\mathbf{w}) = \mathbb{E}_{(\mathbf{x},y) \sim P} [(y - \mathbf{w}^\mathsf{T} \phi(\mathbf{x}))^2]$. Taking the gradient with respect to $\mathbf{w}$ yields:
\begin{equation}
\nabla \mathcal{L}(\mathbf{w}) = -2 \mathbb{E}[y \phi(\mathbf{x})] + 2 \mathbb{E}[\phi(\mathbf{x}) \phi(\mathbf{x})^\mathsf{T}] \mathbf{w}.
\end{equation}
Taking the second derivative gives:
\begin{equation}
\nabla^2 \mathcal{L}(\mathbf{w}) = 2 \mathbb{E}[\phi(\mathbf{x}) \phi(\mathbf{x})^\mathsf{T}] = 2\Sigma.
\end{equation}
Thus, the Hessian matrix is $2\Sigma$, and therefore $\Sigma_{\text{rest}}', \Sigma_{\text{unlearn}}', \Sigma_{\text{rest}}$ are all positive definite matrices. Furthermore, for computational convenience, we assume whitening is applied to the feature extraction layer outputs, resulting in $\tilde{\phi}(\mathbf{x})$. Consequently,
\begin{equation}
\Sigma_{\text{rest}} = \Sigma_{\text{rest}}' = \Sigma_{\text{unlearn}}' = I
\label{eq:9}
\end{equation}
where $I$ is the identity matrix.

Let the classifier be $f_{\mathbf{w}}(\mathbf{x}) = \mathbf{w}^\mathsf{T} \tilde{\phi}(\mathbf{x})$, the squared loss be $\ell(y, \mathbf{w}^\mathsf{T} \tilde{\phi}(\mathbf{x})) = (y - \mathbf{w}^\mathsf{T} \tilde{\phi}(\mathbf{x}))^2$, and define the expected risks $\mathcal{L}_{\text{rest}}'(\mathbf{w}) = \mathbb{E}_{(\mathbf{x},y) \sim P_{\text{rest}}'}[(y - \mathbf{w}^\mathsf{T} \tilde{\phi}(\mathbf{x}))^2]$ and $\mathcal{L}_{\text{unlearn}}'(\mathbf{w}) = \mathbb{E}_{(\mathbf{x},y) \sim P_{\text{unlearn}}'}[(y - \mathbf{w}^\mathsf{T} \tilde{\phi}(\mathbf{x}))^2]$.

The first-order optimality condition gives 
$
\nabla \mathcal{L}_{\text{rest}}'(\mathbf{w}) = -2 \mathbb{E}_{(\mathbf{x},y) \sim P_{\text{rest}}'}[y \tilde{\phi}(\mathbf{x})] + 2 \mathbb{E}_{(\mathbf{x},y) \sim P_{\text{rest}}'}[\tilde{\phi}(\mathbf{x}) \tilde{\phi}(\mathbf{x})^\mathsf{T}] \mathbf{w}_{\text{rest}}' = 0.
$
Combining with Equation~(\ref{eq:9}) yields:
\begin{equation}
\mathbf{w}_{\text{rest}}' ={\Sigma_{\text{rest}}'}^{-1} \mathbb{E}_{(\mathbf{x},y) \sim P_{\text{rest}}'}[y \tilde{\phi}(\mathbf{x})] = {\mu_{\text{rest}}'}^{(y)}
\label{eq:10}
\end{equation}
Similarly,
\begin{align}
\mathbf{w}_{\text{unlearn}}' &= {\Sigma_{\text{unlearn}}'}^{-1} \mathbb{E}_{(\mathbf{x},y) \sim P_{\text{unlearn}}'}[y \tilde{\phi}(\mathbf{x})] = {\mu_{\text{unlearn}}'}^{(y)}  \label{eq:11} \\
\mathbf{w}_{\text{rest}} &= \Sigma_{\text{rest}} \mathbb{E}_{(\mathbf{x},y) \sim P_{\text{rest}}}[y \tilde{\phi}(\mathbf{x})] = {\mu_{\text{rest}}}^{(y)}  \label{eq:12}
\end{align}

Since $D_{\text{rest}}^{(\text{other})}$, $D_{\text{rest}}$, and $D_{\text{unlearn}}$ are three completely distinct classes, we assume the inter-class label features of $D_{\text{rest}}$ and $D_{\text{unlearn}}$ are orthogonal. Thus,
\begin{gather}
{({\mu_{\text{rest}}'}^{(y)})}^\mathsf{T} {\mu_{\text{unlearn}}'}^{(y)} = 0  \\
{({\mu_{\text{rest}}}^{(y)})}^\mathsf{T} {\mu_{\text{unlearn}}'}^{(y)} = 0  \label{eq:14}
\end{gather}
Given that $D_{\text{rest}} \sim P_{\text{rest}}$ and $D_{\text{rest}}' \sim P_{\text{rest}}$,
\begin{align}
{\mu_{\text{rest}}}^{(y)} &= \mathbb{E}_{\mathbf{x} \sim P_{\text{rest}} \mid \mathbf{x} \in D_{\text{rest}}}\left[y \tilde{\phi}(\mathbf{x})\right] \nonumber \\
&\approx \mathbb{E}_{\mathbf{x} \sim P_{\text{rest}} \mid \mathbf{x} \in D_{\text{rest}}'}\left[(y \tilde{\phi})(\mathbf{x})\right] \nonumber  \\ 
&= {\mu_{\text{rest}}'}^{(y)} \label{eq:15}
\end{align}
Therefore,
\begin{equation}
{({\mu_{\text{rest}}}^{(y)})}^\mathsf{T} {\mu_{\text{rest}}'}^{(y)} \approx {({\mu_{\text{rest}}}^{(y)})}^\mathsf{T} {\mu_{\text{rest}}}^{(y)} > 0                        \label{eq:16}
\end{equation}

Next, we calculate the dot product between $\mathbf{w}_{\text{rest}}$ and $\mathbf{w}_{\text{rest}}'$. Combining with Equation~(\ref{eq:16}) gives:
\[
\langle \mathbf{w}_{\text{rest}}, \mathbf{w}_{\text{rest}}' \rangle = ({\mu_{\text{rest}}}^{(y)})^\mathsf{T} {\mu_{\text{rest}}'}^{(y)} \approx ({\mu_{\text{rest}}}^{(y)})^\mathsf{T} {\mu_{\text{rest}}}^{(y)} > 0
\]
Combining with Equations ~(\ref{eq:14}) and~(\ref{eq:16}), the dot product between $\mathbf{w}_{\text{rest}}$ and $\mathbf{w}_{\text{unlearn}}'$ is:
\begin{equation}
\langle \mathbf{w}_{\text{rest}}, \mathbf{w}_{\text{unlearn}}' \rangle = ({\mu_{\text{rest}}}^{(y)})^\mathsf{T} {\mu_{\text{unlearn}}'}^{(y)} = 0
\end{equation}
Therefore,
\[
\langle \mathbf{w}_{\text{rest}}, \mathbf{w}_{\text{rest}}' \rangle \ge \langle \mathbf{w}_{\text{rest}}, \mathbf{w}_{\text{unlearn}}' \rangle.
\]

The inner product $\langle \mathbf{w}_{\text{rest}}$,$\mathbf{w}_{\text{rest}}' \rangle$ represents an unnormalized cosine similarity measure.  This result demonstrates that, under the stated assumptions, the parameter similarity quantified by the dot product between the model trained on the full training set and a sub-model trained on a subset derived from the same distribution exceeds the similarity with parameters of a model trained on an entirely distinct and independent dataset.

Building upon the relationships derived above, we propose the attack framework illustrated in Figure \ref{fig:param_attack_flowchart}. The first step involves training the models $M_{\text{rest}}^{(1)}$, $ M_{\text{rest}}^{(2)}$, $\cdots$, $M_{\text{rest}}^{(n)}$ on the data subsets $D_{\text{rest}}^{(1)}$, $D_{\text{rest}}^{(2)}$, $ \cdots $, $D_{\text{rest}}^{(n)}$, and the models $M_{\text{unlearn}}^{(1)} $, $M_{\text{unlearn}}^{(2)}$, $ \cdots$, $ M_{\text{unlearn}}^{(m)}$ on $D_{\text{unlearn}}^{(1)}$, $D_{\text{unlearn}}^{(2)}$, $ \cdots$, $ D_{\text{unlearn}}^{(m)}$, respectively. Subsequently, we calculate the dot product between the fully connected layer parameters of each trained model, denoted as $w_{\text{rest}}^{(i)}$ and $w_{\text{unlearn}}^{(j)}$ and the target parameters $w_{\text{rest}}$ from the server's unlearned model $M_{\text{rest}}$.

These dot products are used to construct a new dataset for classification:
\[
\mathcal{D}_{\text{dot}} = \left\{ \left( \langle w_{\text{rest}}, w_{\text{rest}}^{(i)} \rangle, \, 0 \right)_{i=1}^{n} \, \cup \, \left( \langle w_{\text{rest}}, w_{\text{unlearn}}^{(j)} \rangle, \, 1 \right)_{j=1}^{m} \right\},
\]
where the label $0$ indicates a non-target  class for inference, and the label $1$ indicates the target  class for inference. Finally, a classification screening is performed on this dataset. The forgotten class can be determined either by deriving an optimal threshold using Youden’s Index or by obtaining a decision boundary through the k-means clustering algorithm.

\subsubsection{Attack Method based on Model Parameter Difference}
Beyond computing the dot product, we also demonstrate that characteristic differences exist between the parameter difference vectors $\mathbf{w}_{\text{rest}} - \mathbf{w}_{\text{rest}}'$ and $\mathbf{w}_{\text{rest}} - \mathbf{w}_{\text{unlearn}}'$, which can be exploited for inference attacks. Combining Equations~(\ref{eq:10}),~(\ref{eq:11}), and~(\ref{eq:12}) yields:
\begin{align*}
\mathbf{w}_{\text{rest}} - \mathbf{w}_{\text{rest}}' &= {\mu_{\text{rest}}}^{(y)} - {\mu_{\text{rest}}'}^{(y)} \\
\mathbf{w}_{\text{rest}} - \mathbf{w}_{\text{unlearn}}' &= {\mu_{\text{rest}}}^{(y)} - {\mu_{\text{unlearn}}'}^{(y)}
\end{align*}
From Equation~(\ref{eq:15}), we have $\mathbf{w}_{\text{rest}} - \mathbf{w}_{\text{rest}}' \rightarrow 0$, while the difference between the distributions $P_{\text{rest}}$ and $P_{\text{unlearn}}'$ implies $\| \mathbf{w}_{\text{rest}} - \mathbf{w}_{\text{unlearn}}' \| > \| \mathbf{w}_{\text{rest}} - \mathbf{w}_{\text{rest}}' \|$. A classifier can therefore distinguish between these two types of difference vectors.

Since we train multiple models, we obtain multiple difference vectors that serve as samples. As illustrated in Figure \ref{fig:param_attack_flowchart}, we construct a new dataset as follows:
\[
\mathcal{D}_{\text{diff}} = \left\{ \left( \mathbf{w}_{\text{rest}}'^{(i)} - \mathbf{w}_{\text{rest}}, \, 0 \right)_{i=1}^{n} \ \cup \ \left( \mathbf{w}_{\text{unlearn}}'^{(j)} - \mathbf{w}_{\text{rest}}, \, 1 \right)_{j=1}^{m} \right\},
\]
where the label '0' corresponds to difference vectors from retained class models, and '1' corresponds to those from forgotten class candidate models. For the classification of this dataset, we employ a Decision Tree classifier. Decision Trees offer practical advantages in computational efficiency and ease of implementation. They can directly handle high-dimensional continuous features without requiring strict parameter normalization. Furthermore, by recursively partitioning the feature space, they can automatically capture non-linear boundaries inherent in the parameter update patterns. Training a Decision Tree on $\mathcal{D}_{\text{diff}}$ enables effective discrimination between the classes corresponding to the retained data and the forgotten data candidates.

\subsection{Label Inference Attack via Model Inversion}

\begin{figure}[htbp]
    \centering
    \includegraphics[width=\linewidth]{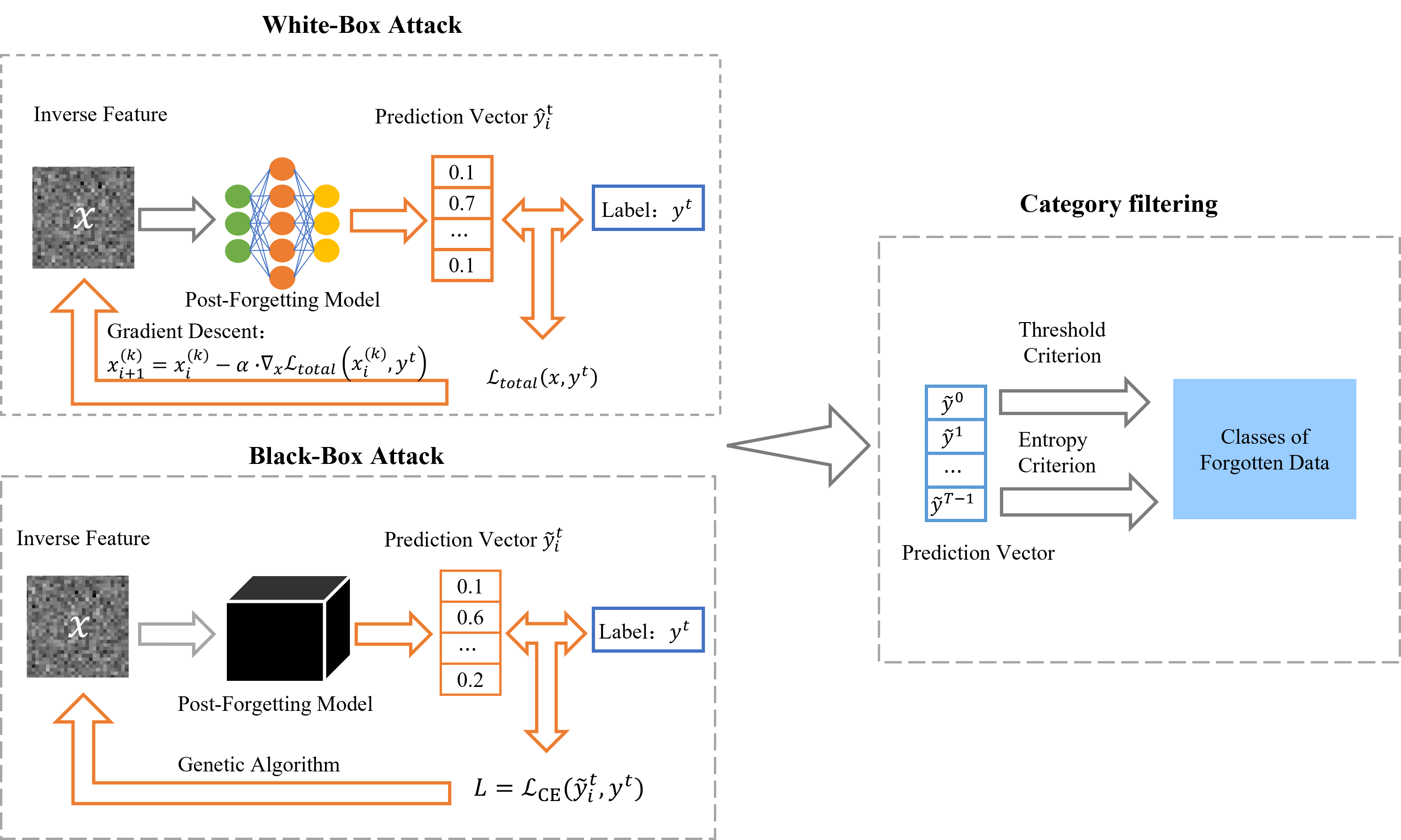} 
    \caption{Framework of the Label Inference Attack via Model Inversion}
    \label{fig:inversion_framework}
\end{figure}

This section first introduces the principle of model inversion attacks. It then describes how an attacker obtains prediction vectors for various classes of training data via model inversion in both white-box and black-box attack scenarios. Finally, it explains how the threshold criterion and entropy criterion are used to identify the forgotten class from the collected prediction vectors. Figure \ref{fig:inversion_framework} illustrates the framework of the attack model.

\subsubsection{Model Inversion Attack}

Let the target model be $M: \mathbb{R}^d \rightarrow \mathbb{R}^T$, where $d$ denotes the dimensionality of the input space. For image data, $d = C \times H \times W$, corresponding to the number of channels, height, and width, respectively. $T$ is the total number of classes.

In a model inversion attack, the features of a specific class from the training data are reconstructed iteratively by minimizing a loss function. This optimization process can be formalized as:
\begin{equation}
    \mathbf{x}_{i+1} = \mathbf{x}_i - \alpha \frac{\partial L(M(\mathbf{x}_i), y_{\text{target}})}{\partial \mathbf{x}} \label{eq:18}
\end{equation}
where $\mathbf{x}_i \in \mathbb{R}^d$ is the sample to be recovered. Its initial value is not drawn from real data but generated from a random distribution. $\alpha$ is the learning rate, $y_{\text{target}}$ is the target class label, and $L$ is the loss function.

\begin{figure}[htbp]
    \centering
    \includegraphics[width=0.23\linewidth]{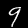}\hfill
    \includegraphics[width=0.23\linewidth]{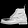}\hfill
    \includegraphics[width=0.23\linewidth]{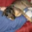}\hfill
    \includegraphics[width=0.23\linewidth]{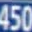}
    
    \vspace{0.5cm} 
    
    \includegraphics[width=0.23\linewidth]{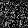}\hfill
    \includegraphics[width=0.23\linewidth]{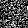}\hfill
    \includegraphics[width=0.23\linewidth]{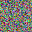}\hfill
    \includegraphics[width=0.23\linewidth]{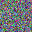}
    \caption{Real Features (top row) versus Inverted Features (bottom row) of Target Classes}
    \label{fig:real_vs_inv_features}
\end{figure}

Intuitively, if a class of data has been successfully forgotten, performing inversion on the post-unlearning model should fail to recover its features; conversely, features should be recoverable. However, the classic gradient inversion method shown in Equation~(\ref{eq:18}) is only effective for linear models or shallow networks. As shown in Figure \ref{fig:real_vs_inv_features}, when applied to DNNs like LeNet or ResNet-18, the inverted images appear as near-random noise, exhibiting a significant visual discrepancy from real samples. This prevents a straightforward visual comparison to determine the forgetting state. Although existing research has introduced Generative Adversarial Networks to enhance inversion for deep networks, these methods heavily rely on prior knowledge of the training data distribution, which contradicts the zero-shot attack scenario assumed in this paper. Therefore, we focus on the stringent condition where the attacker has no access to any information about the training data distribution, exploring the problem of inferring the forgotten class in DNNs.
\begin{figure}[htbp]
    \centering
    \includegraphics[width=0.8\linewidth]{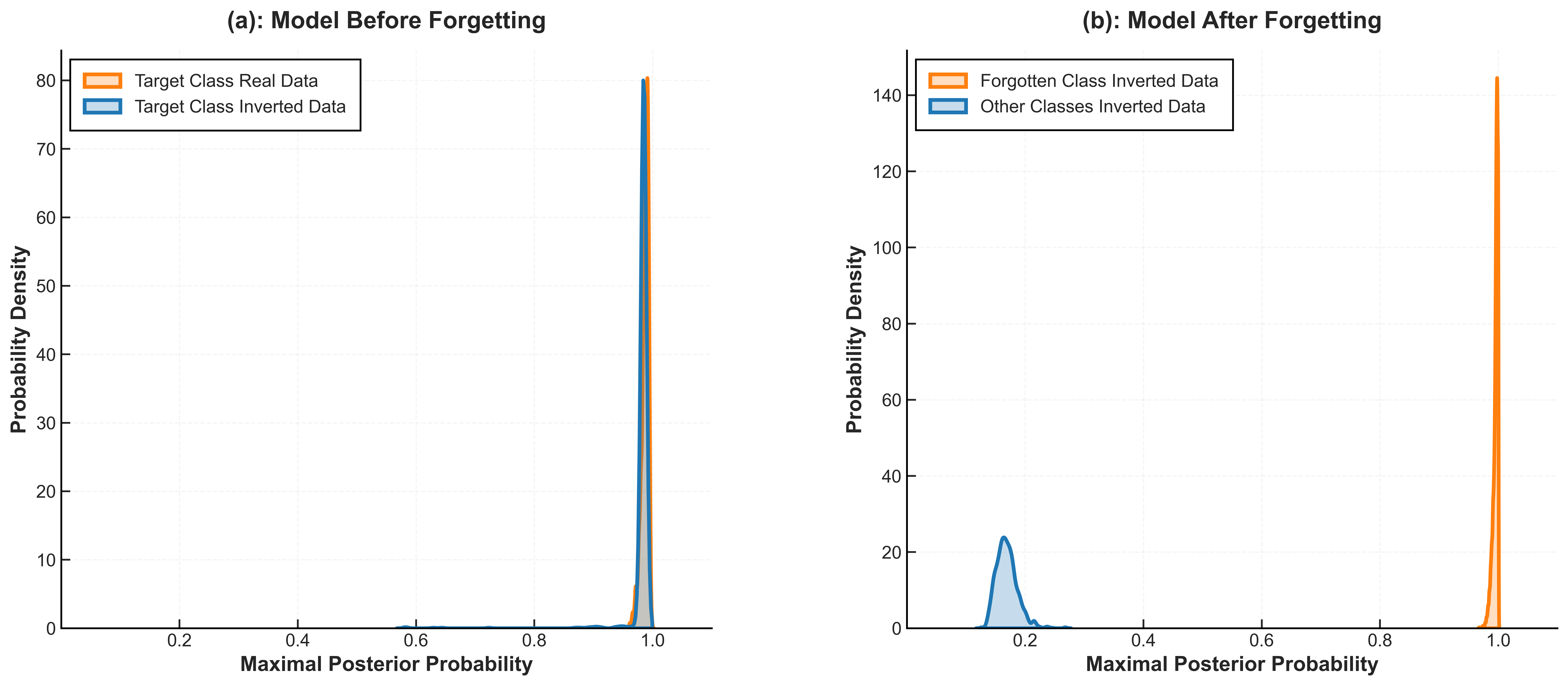} 
    \caption{Prediction Results of the Target Model and the Retrained Model on Inverted Data}
    \label{fig:prediction_comparison}
\end{figure}
Although the inverted samples generated via Equation~(\ref{eq:18}) on DNNs appear as meaningless noise to human visual perception, our experiments reveal that the model still assigns high prediction confidence to these samples. This phenomenon prompts us to re-examine the relationship between inverted samples and model prediction behavior in the context of class-level unlearning. To this end, we conduct inversion attacks on a post-unlearning model based on the retraining mechanism and record the model's \textbf{Maximum Prediction Probability} for the inverted samples. The experimental results are shown in Figure \ref{fig:prediction_comparison}. The analysis indicates: (1) In the pre-unlearning model, the distribution of maximum prediction probabilities for inverted samples of the target class is highly consistent with that of real samples, verifying that the model treats the inversion results as representative of that class despite their lack of visual recognizability. (2) In the post-unlearning model, the discriminant information for non-forgotten classes is preserved, so their inverted samples maintain a high maximum prediction probability. In contrast, the inverted samples of the \textit{forgotten} class exhibit a significant drop in their maximum prediction probability.

The results above demonstrate that the Maximum Prediction Probability can serve as a robust indicator to distinguish between forgotten and non-forgotten classes. The inverted samples of the forgotten class are typically accompanied by low prediction confidence, while those of the retained classes maintain high confidence. Consequently, even though Equation~(\ref{eq:18}) fails at the visual level, this paper can still construct effective attacks for inferring the forgotten class under both white-box and black-box settings, based on the model's prediction probabilities for the inverted samples.

\subsubsection{White-box Attack}

In the white-box attack setting, the attacker has full access to the architecture and parameters of the target model $M$ and can compute the gradients of the model's output with respect to its input. Given a classification task with $T$ classes, our objective is to, for each class label $y^t$ where $t \in \{0, 1, 2, \ldots, T-1\}$, generate an inverted sample that maximizes the model's confidence for that class. This is achieved by optimizing a variable $\mathbf{x}$ in the input space, thereby characterizing the model's representational capacity for classes after the unlearning operation.

Since this optimization problem is non-convex, a single initialization is prone to convergence to a local optimum. Therefore, we introduce a multi-initialization strategy. For each class $y^t$, multiple initial points are sampled independently from a standard Gaussian distribution:
\begin{equation}
    \mathbf{x}_0^{(k)} \sim \mathcal{N}(0, I), \quad k = 1, 2, \cdots, K
\end{equation}
where the superscript $(k)$ denotes the $k$-th independent initialization, and the subscript $0$ denotes the initial step of the optimization process.

For each initialization trajectory, the input is optimized iteratively by minimizing the following objective function:
\begin{equation}
    \min_{\mathbf{x}} \mathcal{L}_{\text{total}}(\mathbf{x}, y^t) = \mathcal{L}_{\text{CE}}(M(\mathbf{x}), y^t) + \lambda_{\text{L2}} \|\mathbf{x}\|_2^2 + \lambda_{\text{TV}} \cdot \mathcal{L}_{\text{TV}}(\mathbf{x})
\end{equation}
where $\mathcal{L}_{\text{CE}}$ is the cross-entropy loss, $\|\mathbf{x}\|_2^2$ is the L2 regularization term, and $\mathcal{L}_{\text{TV}}(\mathbf{x})$ is the Total Variation (TV) regularization term used to constrain the spatial smoothness of the input. Accordingly, the update process at the $i$-th iteration is expressed as:
\begin{equation}
    \mathbf{x}_{i+1}^{(k)} = \mathbf{x}_i^{(k)} - \alpha \cdot \nabla_{\mathbf{x}} \mathcal{L}_{\text{total}}(\mathbf{x}_i^{(k)}, y^t) \tag{11}
\end{equation}
where $\alpha$ is the learning rate. Upon completion of optimization for each initialization, a candidate solution $\mathbf{x}_E^{(k)}$ is obtained. Since different initializations may converge to different local optima, we further select the best solution from all candidates:
\begin{equation}
    \mathbf{x}^{*} = \arg\min_{k \in \{1, 2, \cdots, K\}} \mathcal{L}_{\text{CE}}(M(\mathbf{x}_E^{(k)}), y^t).
\end{equation}
Finally, the optimal input $\mathbf{x}^{*}$ is fed into the model to obtain the corresponding prediction probability vector $\tilde{\mathbf{y}}^t = \text{softmax}(M(\mathbf{x}^{*})) \in \mathbb{R}^T$. Repeating the above process for all classes yields the Inverted Prediction Vector (IPV) set: $\mathcal{Y}^{\text{IPV}} = \{\tilde{\mathbf{y}}^0$, $\tilde{\mathbf{y}}^2$, $\cdots, \tilde{\mathbf{y}}^{T-1} \}$.

\subsubsection{Black-box Attack}

In the black-box attack setting, the attacker cannot access the internal architecture or parameters of the target model $M$ and cannot utilize gradient information to optimize the input. Consequently, the attacker can only iteratively adjust the input sample based on the feedback signal from the model's output. To address this, we formulate the generation of adversarial samples as a population-based evolutionary optimization process, employing an improved genetic algorithm to search the input space for samples that induce the model to output the target class.

We normalize the input $\mathbf{x}$ such that $\mathbf{x} \in [0, 1]^d$. For a given target class $y^t$, the attack objective is to find the optimal input $\mathbf{x}^*$ within this space that maximizes the model's predicted probability for the target class:
\begin{equation}
    \mathbf{x}^* = \arg\max_{\mathbf{x} \in [0, 1]^d} P(y^t | \mathbf{x})
\end{equation}
where $P(y^t | \mathbf{x})$ denotes the target class probability output by the softmax layer of model $M$. Based on this objective, we directly define the target class probability as the fitness function:
\begin{equation}
    F_{\text{fitness}}(\mathbf{x}) = P(y^t | \mathbf{x}).
\end{equation}

To achieve this objective, we first construct the initial population $\mathcal{P}^{(0)} = \{\mathbf{x}_1$, $\mathbf{x}_2$, $\cdots$, $\mathbf{x}_N\}$ by random sampling from a uniform distribution over the input space. Subsequently, in each generation $g$ of the iterative process, we evaluate the fitness of all individuals in the population. Based on the fitness values, a selection operation is performed to choose parent pairs from the current population according to a probability distribution. For a selected pair of parent individuals $\mathbf{x}^{(p_1)}$ and $\mathbf{x}^{(p_2)}$, a single-point crossover strategy is employed to generate an offspring individual. This involves concatenating the gene sequences at a randomly chosen position $k \in \{1$, $\cdots$, $d\}$, yielding the offspring:
\begin{equation}
    \mathbf{x}^{(c)} = [\mathbf{x}^{(p_1)}_{1:k}, \mathbf{x}^{(p_2)}_{k+1:d}].
\end{equation}

Following the crossover operation, a random mutation is applied to the offspring to enhance population diversity and avoid convergence to local optima. A random dimension $j$ of its chromosome is selected, and a random perturbation $\epsilon \sim \mathcal{N}(0, 1)$ is introduced:
\begin{equation}
    \mathbf{x}^{(c)}_j = \mathbf{x}^{(c)}_j + \epsilon.
\end{equation}
A projection operation is then applied to ensure the mutated individual remains within the valid input space $[0, 1]^d$.

To balance global exploration and local exploitation, we introduce an adaptive mutation mechanism where the mutation strength decreases gradually over generations:
\begin{equation}
    \sigma^{(g+1)} = \alpha \cdot \sigma^{(g)}, \quad \alpha \in (0, 1)
\end{equation}
where $\alpha$ is the decay factor. This strategy allows the algorithm to maintain strong exploratory capability in the early stages while gradually focusing on high-fitness regions in later stages, thereby improving convergence stability.

Furthermore, to prevent the loss of high-quality individuals during evolution, we incorporate an elite preservation strategy. In each generation, the top $k$ individuals with the highest fitness are preserved and directly transferred to the next generation's population, ensuring monotonic improvement of the best-found solution.

After $G$ generations of iteration, the individual with the highest fitness is selected from the final population as the optimal solution $\mathbf{x}^*$. This solution is then fed into the target model to obtain the corresponding prediction output $\tilde{\mathbf{y}}^t = \text{softmax}(M(\mathbf{x}^*))$. Repeating the above process for all target classes $y^t$ yields the Inverted Prediction Vector  set $\mathcal{Y}^{\text{IPV}}$.

\subsubsection{Class Screening}

After obtaining the inverted prediction vectors $\mathcal{Y}_{\text{IPV}}$ for all classes, the forgotten classes must be identified from $\mathcal{Y}^{\text{IPV}}$. We propose two screening criteria: the Threshold Criterion and the Entropy Criterion.

\textbf{Threshold Criterion.} The core principle is that if the maximum prediction probability $\tilde{y}^t$ for the inverted data of the $i$-th class falls below a predefined threshold, that class is considered forgotten. A dynamic threshold selection method is designed as follows. First, we compute the mean and standard deviation of the maximum prediction probabilities:
\begin{align}
    \mu_{\text{max}} &= \frac{1}{T} \sum_{i=0}^{T-1} \tilde{y}^i, \\
    \sigma_{\text{max}} &= \sqrt{ \frac{1}{T} \sum_{i=0}^{T-1} (\tilde{y}^i - \mu_{\text{max}})^2 }.
\end{align}
The dynamic threshold $\theta$ is then defined as:
\begin{equation}
    \theta = \mu_{\text{max}} - \alpha \cdot \sigma_{\text{max}}
\end{equation}
where $\alpha$ is a tunable coefficient. The index set of forgotten classes is:
\begin{equation}
    \mathcal{T}_{\text{forgotten}} = \{ i \mid \tilde{y}^i < \theta \}.
\end{equation}

\textbf{Entropy Criterion.} For data points that the model has previously encountered, the predictions demonstrate elevated confidence characterized by minimal entropy; in contrast, for data that remain unseen or have been subjected to unlearning, the predictions exhibit diminished confidence accompanied by heightened entropy.To mitigate class order effects, each prediction vector in $\mathcal{Y}_{\text{IPV}}$ is sorted by probability values, yielding $\widetilde{\mathcal{Y}}^{\text{IPV}}$. K-means clustering ($K=2$) partitions $\widetilde{\mathcal{Y}}^{\text{IPV}}$ into two clusters with index sets $\text{index}_1$ and $\text{index}_2$, corresponding to subsets $\mathcal{Y}^1$ and $\mathcal{Y}^2$ of the original vectors. The average information entropy for subset $\mathcal{Y}^k$ is:
\begin{equation}
    \bar{H}^k = \frac{1}{|\mathcal{Y}^k|} \sum_{\mathbf{p} \in \mathcal{Y}^k} H(\mathbf{p}), \quad H(\mathbf{p}) = -\sum_{j=1}^{T} p_j \log p_j, \quad k \in \{1, 2\}.
\end{equation}
The cluster with higher average entropy is identified as containing the forgotten classes. The class index corresponding to the maximum probability in each vector of this cluster constitutes the forgotten class set.

\section{Experimental Settings}

\subsection{Datasets and Models}

To comprehensively evaluate the performance of the proposed method, we conducted experiments on four widely used benchmark datasets: MNIST, Fashion-MNIST, SVHN, and CIFAR-10. These datasets are described in detail below.

The\textbf{ SVHN } dataset\cite{netzer2011reading} consists of cropped digit images extracted from Google Street View images. It contains 73,257 training images and 26,032 testing images. Unlike the previously mentioned datasets, SVHN images are 32$\times$32 pixel RGB color images. The backgrounds typically contain complex natural scene noise, and the digits exhibit significant variations in position and scale, making this dataset more representative of real-world application scenarios.

The \textbf{CIFAR-10} dataset \cite{cifar10} is a small-scale color image dataset containing everyday objects. It consists of 60,000 3$\times$32$\times$32 pixel RGB images, partitioned into 50,000 training samples and 10,000 testing samples. The dataset encompasses 10 classes: airplane, automobile, bird, cat, deer, dog, frog, horse, ship, and truck.

The \textbf{MNIST} dataset \cite{mnist} is a classic handwritten digit recognition dataset. It contains 60,000 training images and 10,000 testing images. All samples are single-channel grayscale images with a uniform size of 28$\times$28 pixels. The dataset covers ten digit classes (0 through 9) and is a commonly used benchmark for validating the fundamental performance of image classification algorithms.

The \textbf{Fashion-MNIST} dataset \cite{xiao2017fashion} comprises images of fashion products. It contains 60,000 training samples and 10,000 testing samples, spanning ten clothing and accessory categories: T-shirt, Trouser, Pullover, Dress, Coat, Sandal, Shirt, Sneaker, Bag, and Ankle boot.
\begin{table}[htbp]
\centering
\caption{Mapping between Models and Datasets}
\label{tab:model_dataset_mapping}
\begin{tabular}{cc}
\Xhline{1pt}
Model & Dataset \\
\hline
\multirow{2}{*}{ResNet18} & SVHN \\
& CIFAR10 \\
\hline
\multirow{2}{*}{LeNet} & MNIST \\
& FASHION-MNIST \\
\Xhline{1pt}
\end{tabular}
\end{table}

Based on the characteristics of the above datasets, we respectively selected LeNet and ResNet18 as the base model architectures, and conducted experiments using the corresponding relationships shown in Table~\ref{tab:model_dataset_mapping}.

\subsection{Evaluation Metrics}

In the context of Class-level Unlearning, we formalize the attack objective as a binary classification task: determining whether a given class is a forgotten or mon-forgotten class. We define the Attack Success Rate (ASR) as:

\begin{equation}
\text{ASR} = \frac{N_{\text{correct}}}{N} \times 100\% \tag{12}
\end{equation}

where $N_{\text{correct}}$ denotes the number of correctly inferred classes, and $N$ is the total number of classes. This metric reflects the attacker's global inference accuracy regarding the forgetting behavior across the entire class space. An ASR significantly exceeding the random guessing baseline of 50\% indicates that the unlearned model is vulnerable to privacy leakage via the proposed inference attack.

\subsection{Target Unlearning Algorithms}

To systematically assess the privacy leakage risks in Class-level Unlearning scenarios, this paper selects five representative machine unlearning algorithms as attack targets. These methods span from the most naive retraining strategy to mainstream parameter editing and optimization direction reversal techniques in recent years. The specific algorithms are detailed below.

\textbf{Re-Train (RT).} Re-Training refers to the process of retraining a model from scratch after removing all data designated for forgetting from the training set. Although this method theoretically guarantees optimal unlearning performance, it is typically used only as the performance upper-bound benchmark in machine unlearning research due to its prohibitively high computational cost and inefficiency.

\textbf{Fine-Tune (FT).} Fine-Tuning involves performing a small number of training iterations on the original model using the remaining data after deleting the data to be forgotten, usually with a high learning rate. This method exploits the property of \textit{Catastrophic Forgetting} \cite{french1999catastrophic, goodfellow2013empirical} in neural networks, where learning a new distribution (the remaining data) rapidly overwrites old knowledge (the data to be forgotten).

\textbf{Random Label (RL).} The Random Label method disrupts the model's memory by corrupting the supervision signals of the data to be forgotten. It replaces the labels of the forgetting samples with randomly generated incorrect labels, and then continues training with these mislabeled data. This misleading gradient update aims to confuse the model's decision boundary for the specific class \cite{graves2021amnesiac}.

\textbf{Amnesiac Unlearn (AU).} Amnesiac Unlearning is a parameter-contribution-based method. During the training phase, it tracks and records the update contribution of each sample to the model parameters. When performing the unlearning operation, it directly subtracts the cumulative update amount corresponding to the data to be forgotten from the model parameters, achieving precise erasure \cite{graves2021amnesiac}.

\textbf{Negative Gradient (NG).} The Negative Gradient method formulates unlearning as the inverse process of training. While conventional training aims to minimize the loss function, NG instead updates the model parameters by performing Gradient Ascent on the data to be forgotten, thereby maximizing the loss value for those samples. This forces the model to forget the specific data \cite{liu2022backdoor}.

\section{Experimental Results}
This section first validates the unlearning performance of the selected algorithms on the MNIST, Fashion-MNIST, SVHN, and CIFAR-10 datasets. It then presents the attack results of the proposed methods---based on model parameters and inverted data---in both white-box and black-box settings, for both single-class and multi-class unlearning scenarios. It is noteworthy that in the experiments, the multi-class unlearning scenario involves forgetting three randomly selected classes. Finally, we provide a statistical comparison of the results and analyze the impact of the number of forgotten classes on the Attack Success Rate.

\subsection{Performance of the Various Unlearning Methods}

\begin{table*}[htbp]
\centering
\caption{Performance of Unlearning Algorithms in Single-Class Unlearning Scenarios}
\label{tab:unlearning_performance_single}
\begin{tabular}{l c c c c c}
\Xhline{1pt}
\multirow{3}{*}{\makecell{Forgetting\\Scenario}} & \multirow{3}{*}{\makecell{Forgetting\\Method}} & \multicolumn{2}{c}{$D_f$ (\%)} & \multicolumn{2}{c}{$D_r$ (\%)} \\
\cline{3-6}
& &\makecell{Before\\Forgetting} & \makecell{After\\Forgetting} & \makecell{Before\\Forgetting} & \makecell{After\\Forgetting} \\
\hline
\multirow{5}{*}{MNIST} & RT & 99.11 & 28.30 & 98.99 & 99.06 \\
& FT & 98.85 & 8.60 & 99.01 & 99.16 \\
& RL & 98.51 & 0.00 & 98.82 & 99.01 \\
& AU & 98.97 & 0.06 & 98.89 & 99.03 \\
& NG & 98.94 & 0.00 & 99.00 & 98.21 \\
\hline
\multirow{5}{*}{\makecell{Fashion\\-MNIST}} & RT & 88.92 & 0.15 & 89.88 & 91.42 \\
& FT & 87.70 & 0.02 & 90.01 & 91.22 \\
& RL & 88.77 & 0.00 & 88.18 & 90.31 \\
& AU & 86.82 & 0.00 & 88.61 & 89.88 \\
& NG & 93.10 & 0.00 & 89.41 & 89.33 \\
\hline
\multirow{5}{*}{SVHN} & RT & 91.12 & 0.03 & 92.42 & 93.05 \\
& FT & 91.96 & 11.93 & 92.38 & 92.97 \\
& RL & 91.33 & 0.00 & 92.25 & 92.22 \\
& AU & 88.86 & 0.91 & 92.47 & 92.58 \\
& NG & 92.03 & 28.62 & 92.38 & 91.89 \\
\hline
\multirow{5}{*}{CIFAR10} & RT & 72.84 & 0.08 & 75.29 & 77.60 \\
& FT & 74.45 & 5.29 & 75.11 & 76.95 \\
& RL & 76.15 & 0.00 & 75.03 & 75.37 \\
& AU & 74.13 & 0.94 & 75.84 & 76.31 \\
& NG & 74.10 & 16.89 & 75.14 & 73.43 \\
\Xhline{1pt}
\end{tabular}
\end{table*}

\begin{table*}[htbp]
\centering
\caption{Performance of Unlearning Algorithms in Multi-Class Unlearning Scenarios}
\label{tab:unlearning_performance_multi}
\begin{tabular}{l c c c c c}
\Xhline{1pt}
\multirow{3}{*}{\makecell{Forgetting\\Scenario}} & \multirow{3}{*}{\makecell{Forgetting\\Method}} & \multicolumn{2}{c}{$D_f$ (\%)} & \multicolumn{2}{c}{$D_r$ (\%)} \\
\cline{3-6}
& &\makecell{Before\\Forgetting} & \makecell{After\\Forgetting} & \makecell{Before\\Forgetting} & \makecell{After\\Forgetting} \\
\hline
\multirow{5}{*}{MNIST} & RT & 99.01 & 32.48 & 99.00 & 99.30 \\
& FT & 99.01 & 15.84 & 99.00 & 99.42 \\
& RL & 98.83 & 0.00 & 98.91 & 99.18 \\
& AU & 97.22 & 0.00 & 99.02 & 99.24 \\
& NG & 98.95 & 1.76 & 99.02 & 98.58 \\
\hline
\multirow{5}{*}{\makecell{Fashion\\-MNIST}} & RT & 90.39 & 1.20 & 89.52 & 92.44 \\
& FT & 89.31 & 0.34 & 89.98 & 92.69 \\
& RL & 88.97 & 0.00 & 87.99 & 92.05 \\
& AU & 74.02 & 0.00 & 91.26 & 92.95 \\
& NG & 88.67 & 0.16 & 90.26 & 91.73 \\
\hline
\multirow{5}{*}{SVHN} & RT & 92.48 & 0.39 & 92.23 & 93.90 \\
& FT & 91.97 & 18.46 & 92.40 & 94.10 \\
& RL & 92.10 & 0.00 & 92.19 & 92.93 \\
& AU & 82.73 & 0.53 & 93.04 & 93.22 \\
& NG & 91.63 & 31.53 & 92.55 & 92.83 \\
\hline
\multirow{5}{*}{CIFAR10} & RT & 74.56 & 0.22 & 75.24 & 80.83 \\
& FT & 74.64 & 7.50 & 75.21 & 80.06 \\
& RL & 74.55 & 0.00 & 75.62 & 78.10 \\
& AU & 77.32 & 0.14 & 78.36 & 80.14 \\
& NG & 75.12 & 7.36 & 75.00 & 77.85 \\
\Xhline{1pt}
\end{tabular}
\end{table*}

\begin{figure}[htbp]
\centering
\includegraphics[width=0.4\textwidth]{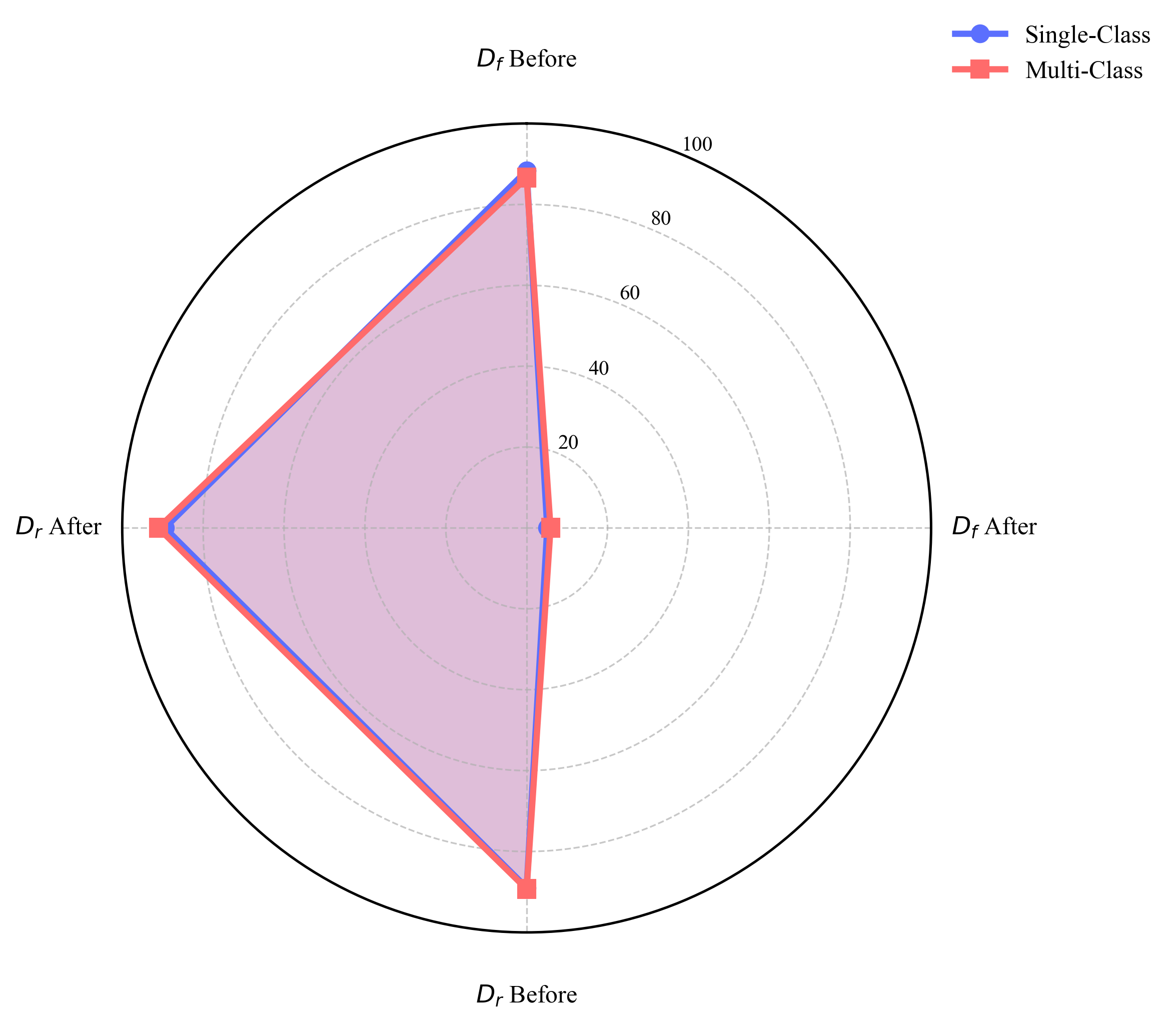} 
\caption{Average Unlearning Performance of the Model under Single-Class and Multi-Class Scenarios}
\label{fig:unlearning_performance_avg}
\end{figure}

This experiment aims to evaluate the unlearning efficacy of multiple model unlearning algorithms in both single-class and multi-class scenarios, and further examine the impact of the unlearning operation on the model's overall performance. Tables~\ref{tab:unlearning_performance_single} and \ref{tab:unlearning_performance_multi} present the experimental results of various unlearning algorithms across different datasets. The evaluation metric is classification accuracy, where $D_f$ and $D_r$ denote the forgotten data and the remaining data, respectively. Figure~\ref{fig:unlearning_performance_avg} shows the average unlearning performance across the different unlearning methods and dataset configurations. The experimental results indicate that, in terms of unlearning performance, all evaluated algorithms achieve a high level of efficacy. The accuracy of the unlearned model on $D_f$ is significantly lower than its accuracy on $D_r$.

\subsection{White-box Attack}

\subsubsection{Unlearning Label Inference from Model Parameters}

\begin{figure}[htbp]
    \centering
    \subfloat[Single-class]{\includegraphics[width=0.45\linewidth]{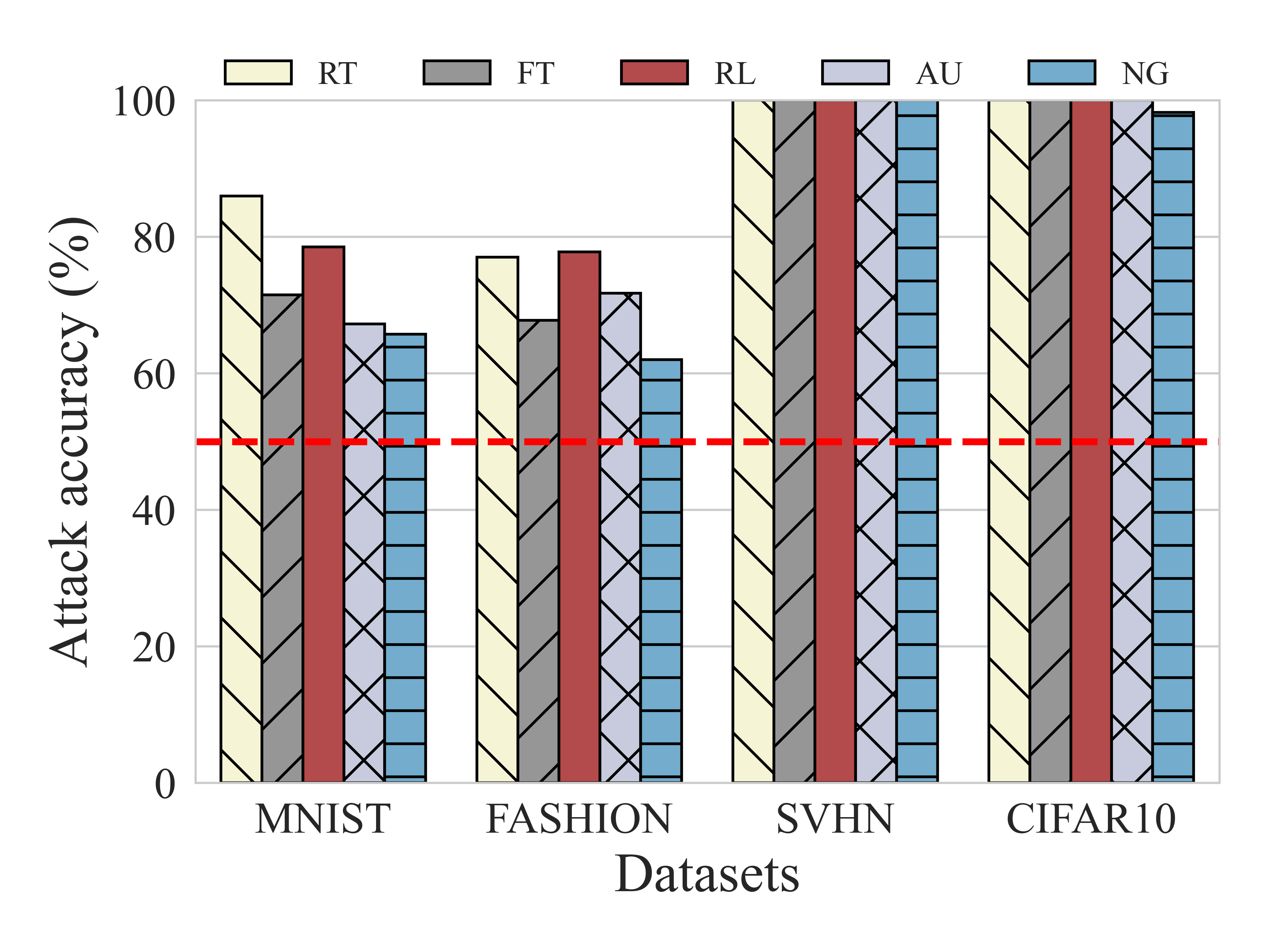}\label{fig:6a}}
    \hfill
    \subfloat[Multi-class]{\includegraphics[width=0.45\linewidth]{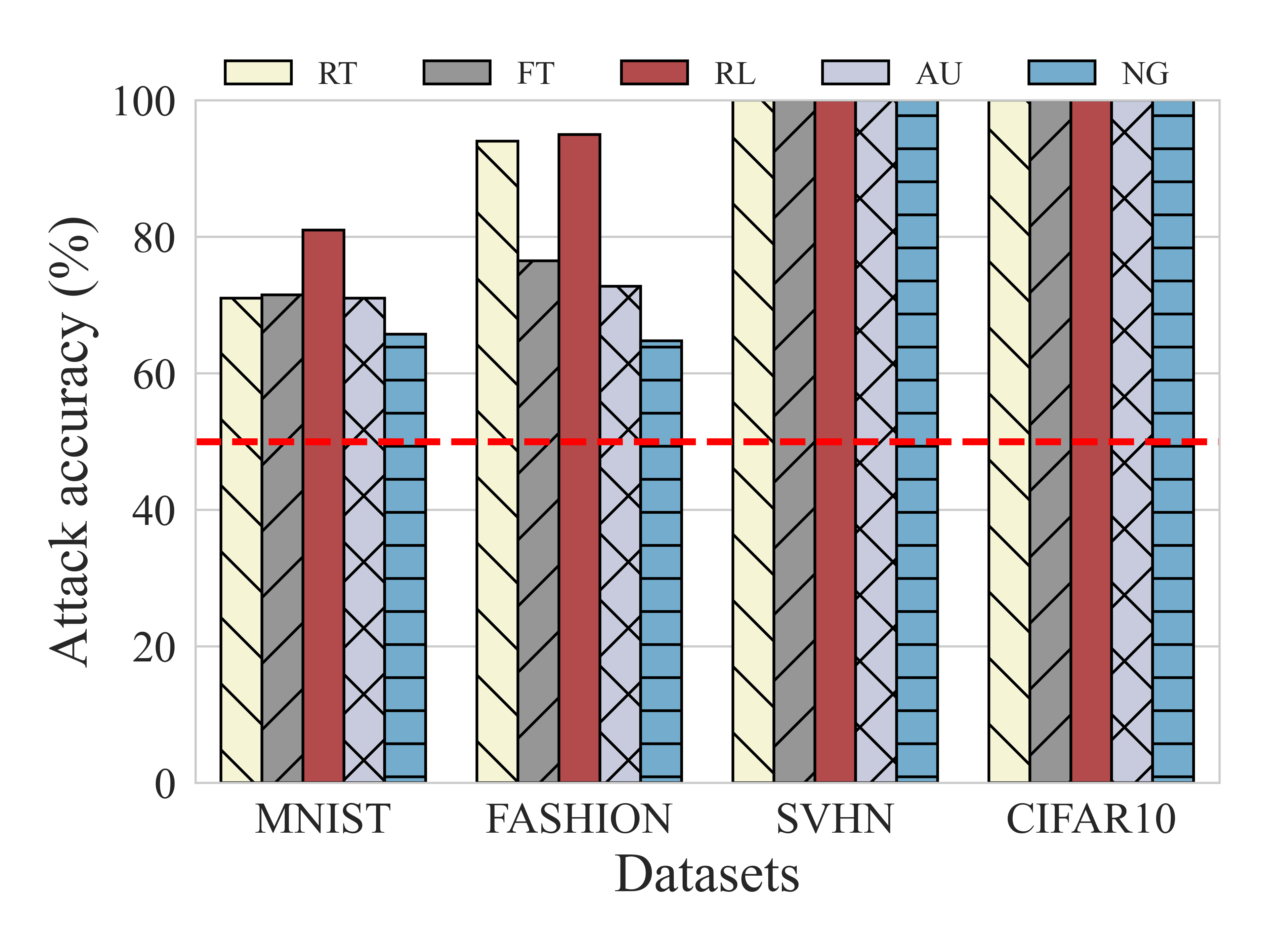}\label{fig:6b}}
    \caption{White-box Attack Based on Youden's Index Threshold and Fully Connected Layer Dot Product}
    \label{fig:youdens_threshold_attack}
\end{figure}

\begin{figure}[htbp]
    \centering
    \subfloat[Single-class]{\includegraphics[width=0.45\linewidth]{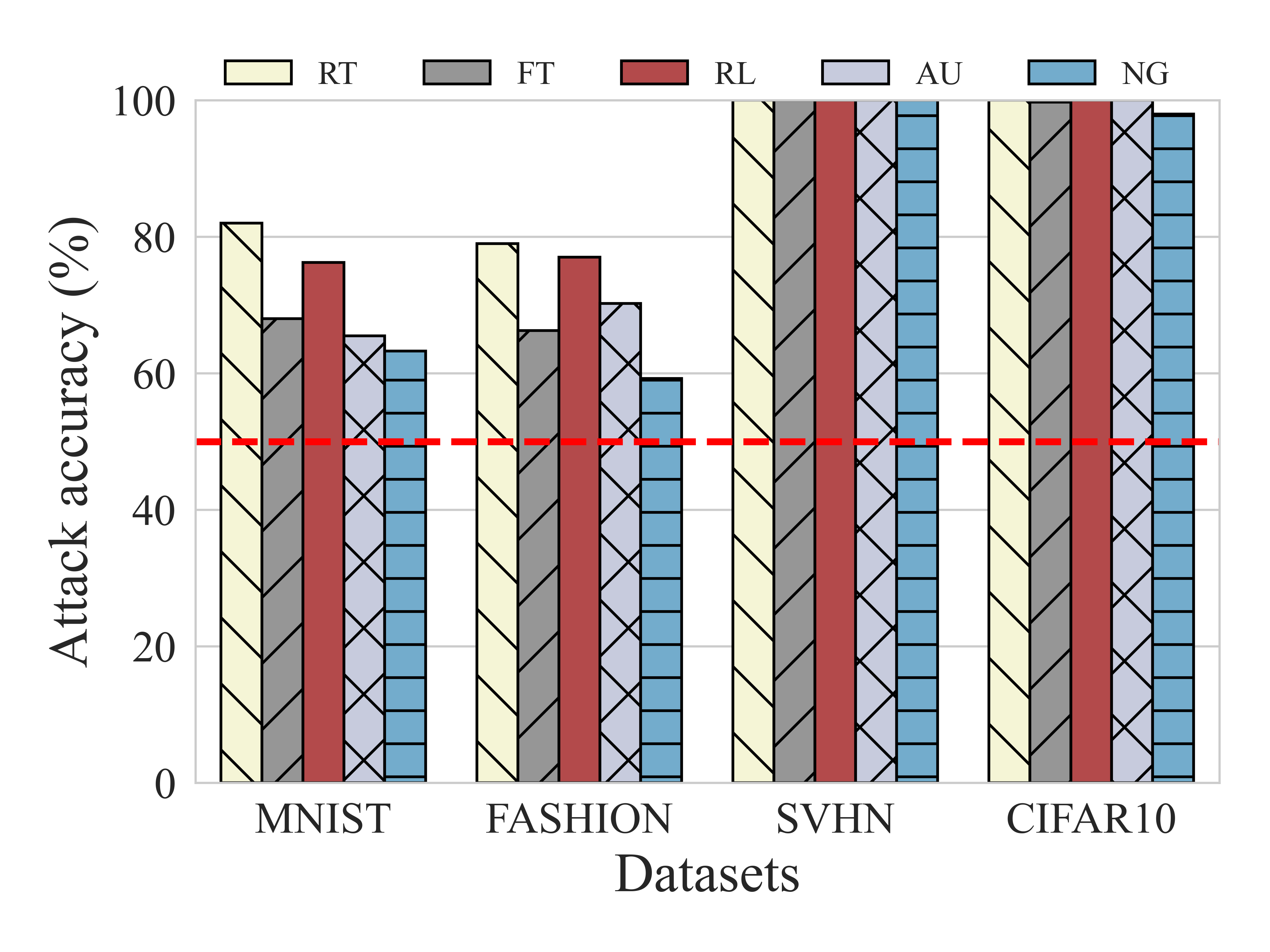}\label{fig:7a}}
    \hfill
    \subfloat[Multi-class]{\includegraphics[width=0.45\linewidth]{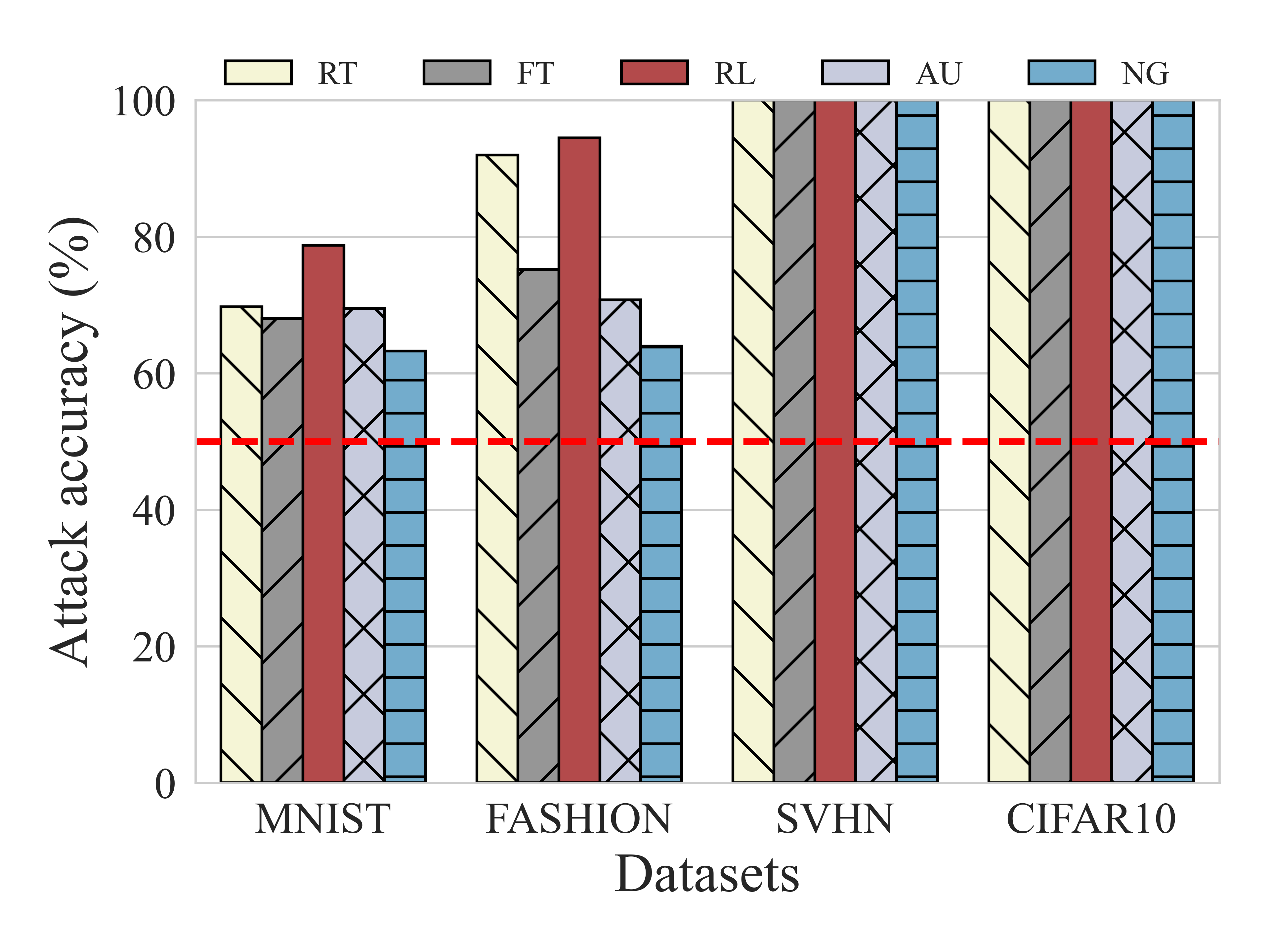}\label{fig:7b}}
    \caption{White-box Attack Based on K-Means and Fully Connected Layer Dot Product}
    \label{fig:kmeans_dot_attack}
\end{figure}

\begin{figure}[htbp]
    \centering
    \subfloat[Single-class]{\includegraphics[width=0.45\linewidth]{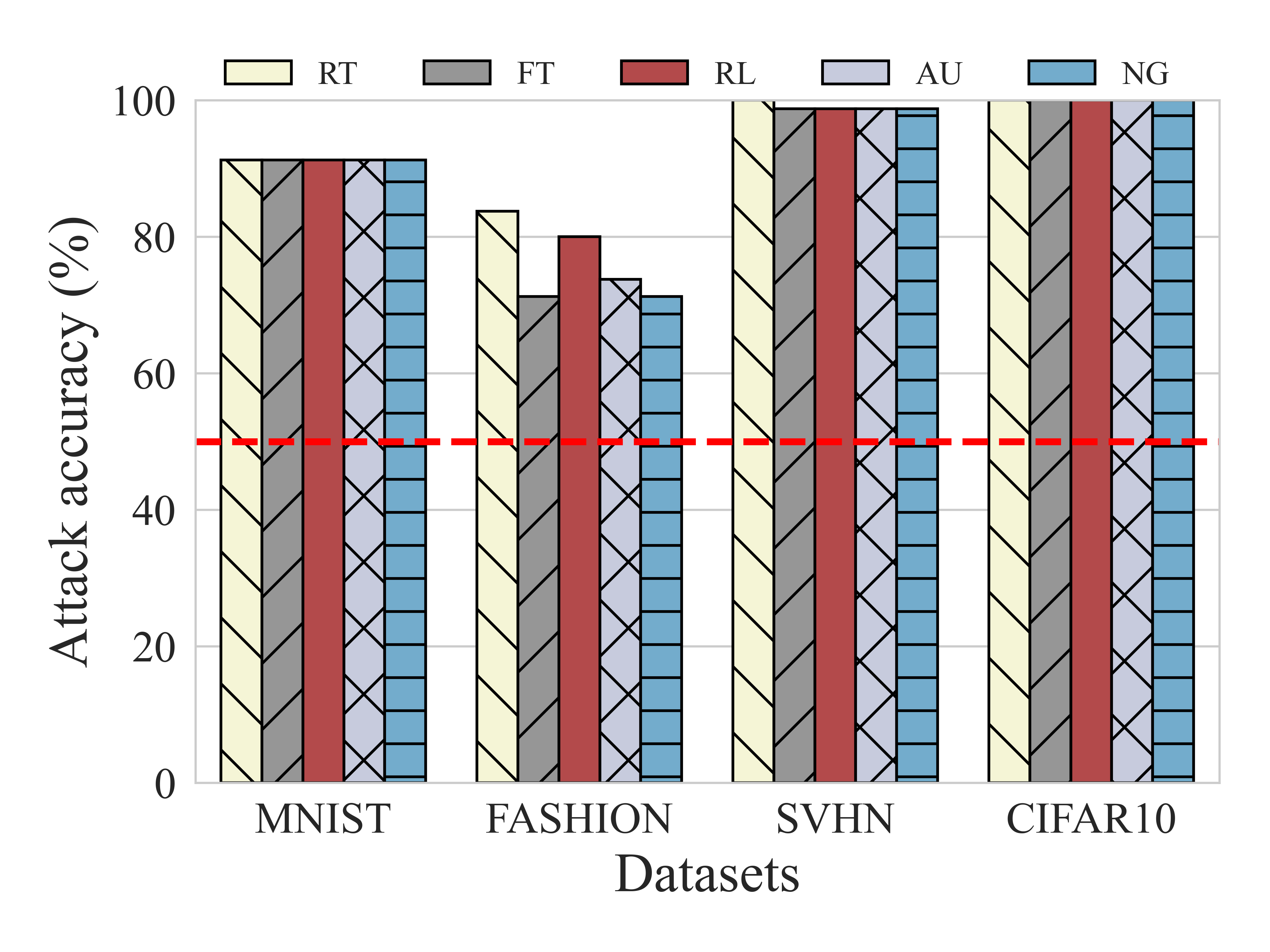}\label{fig:8a}}
    \hfill
    \subfloat[Multi-class]{\includegraphics[width=0.45\linewidth]{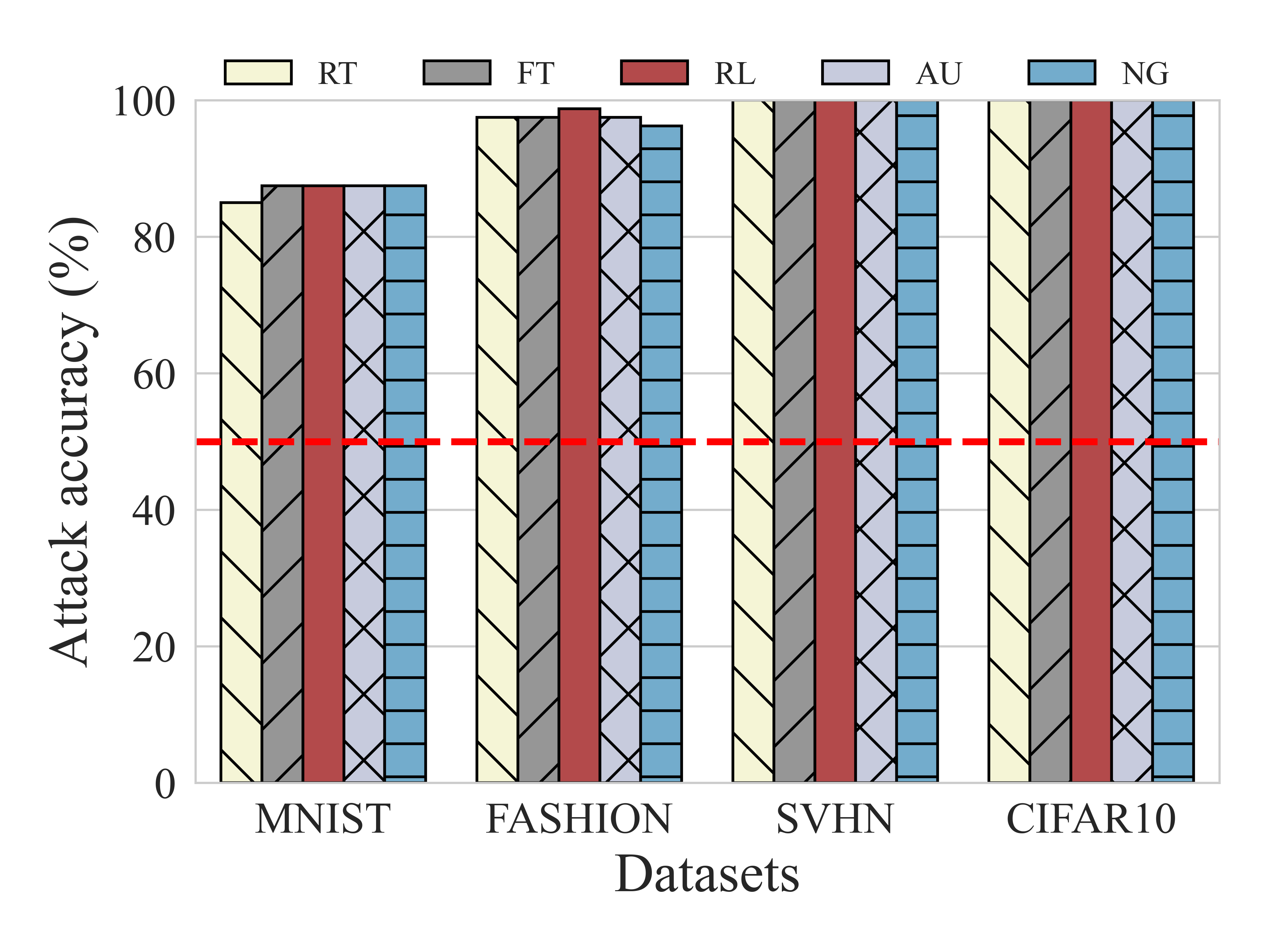}\label{fig:8b}}
    \caption{White-box Attack Based on Decision Tree and Fully Connected Layer Difference}
    \label{fig:dt_diff_attack}
\end{figure}

Under the experimental settings described above, we conduct comprehensive experiments. Figures \ref{fig:youdens_threshold_attack}, \ref{fig:kmeans_dot_attack}, and \ref{fig:dt_diff_attack} present the results of the three proposed white-box attacks based on model parameters. The attack performance on the SVHN and CIFAR-10 datasets is significantly superior to that on the MNIST and Fashion-MNIST datasets, with the ASR approaching 100\% in most scenarios. This discrepancy primarily stems from the fact that the theoretical assumptions underpinning the white-box attack are often difficult to satisfy completely in practical settings. These assumptions include the positive definiteness of the Hessian matrix of the loss function, feature whitening, and the orthogonality of inter-class label vectors. Different models and datasets satisfy these assumptions to varying degrees, resulting in the attack effectiveness being influenced by factors such as model architecture and dataset characteristics.

\begin{figure}[htbp]
    \centering
    \subfloat[Single-class]{\includegraphics[width=0.45\linewidth]{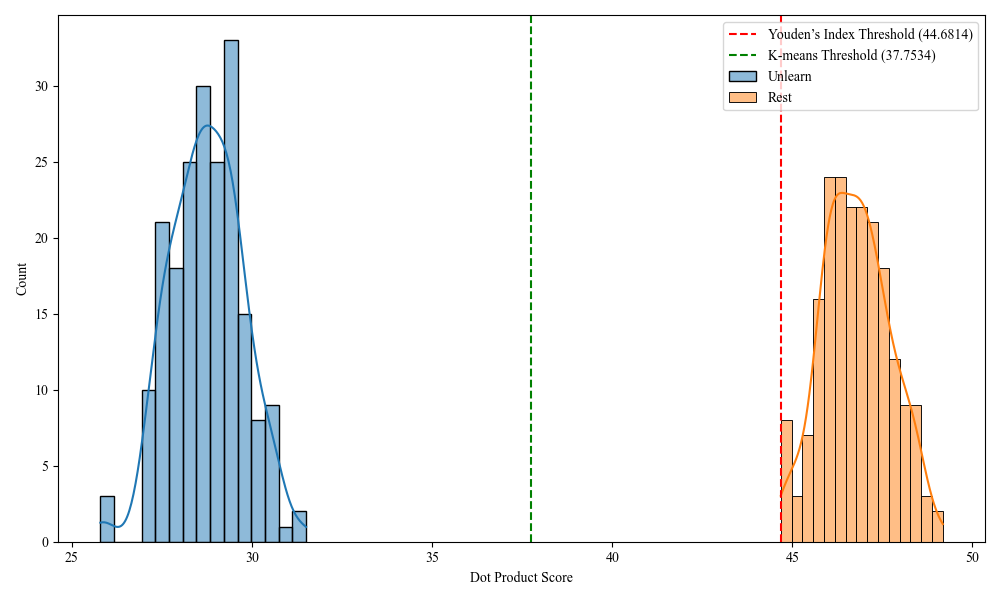}\label{fig:9a}}
    \hfill
    \subfloat[Multi-class]{\includegraphics[width=0.45\linewidth]{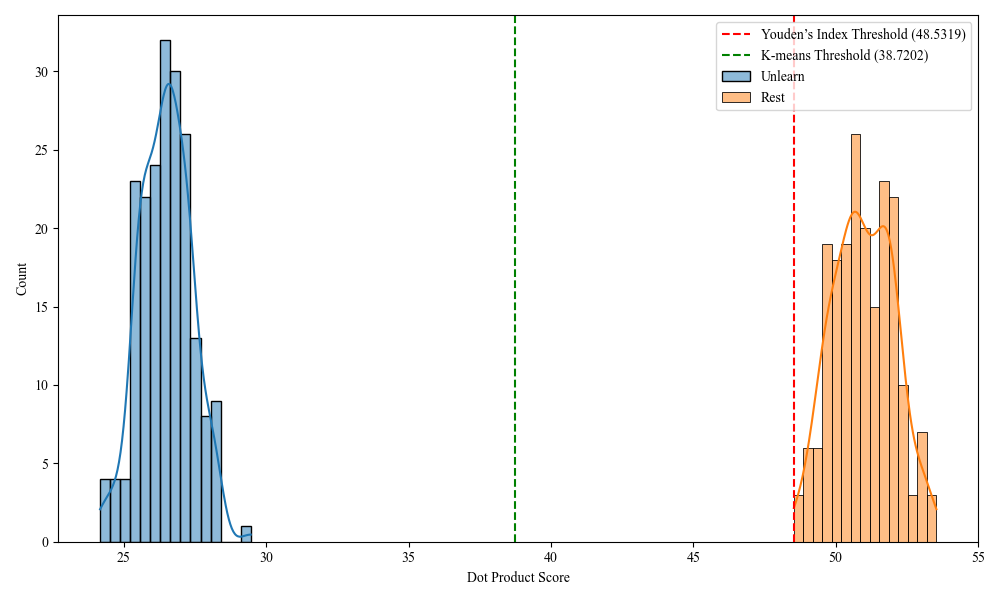}\label{fig:9b}}
    \caption{Distribution of Dot Product Results for the Retrained Model on the CIFAR-10 Dataset}
    \label{fig:cifar10_dot_dist}
\end{figure}

\begin{figure}[htbp]
    \centering
    \subfloat[Single-class]{\includegraphics[width=0.45\linewidth]{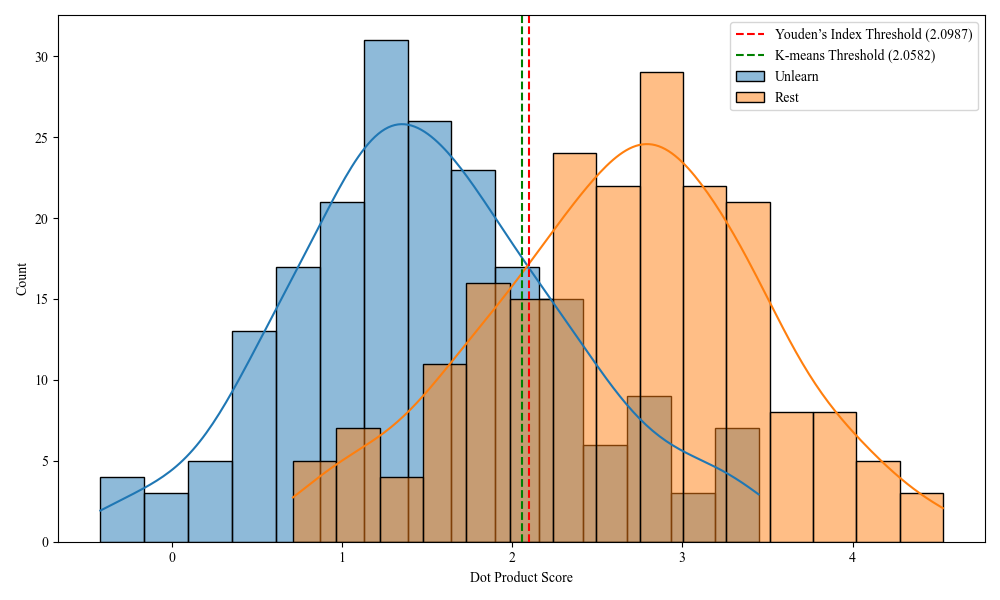}\label{fig:10a}}
    \hfill
    \subfloat[Multi-class]{\includegraphics[width=0.45\linewidth]{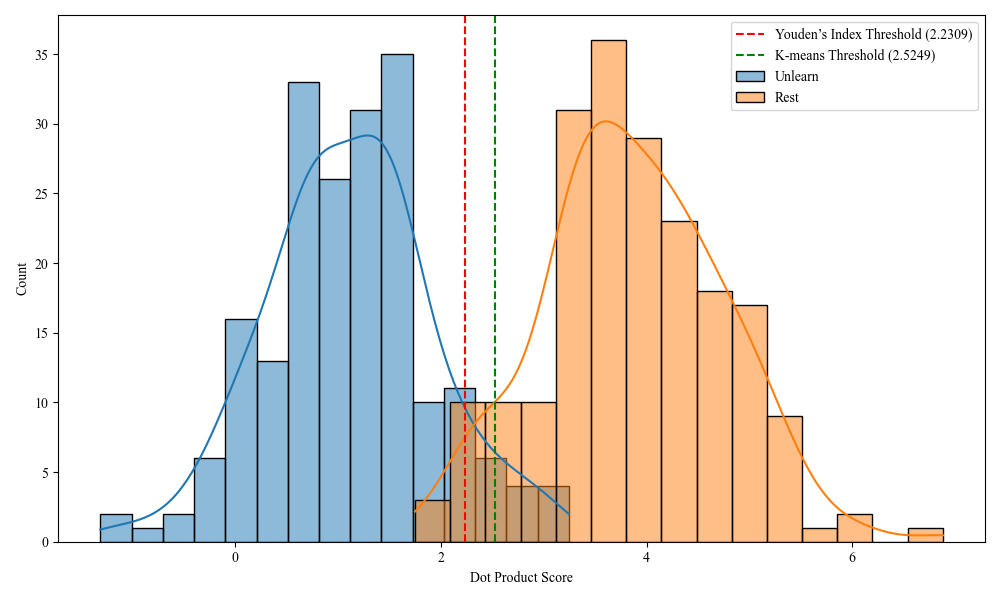}\label{fig:10b}}
    \caption{Distribution of Dot Product Results for the Retrained Model on the Fashion-MNIST Dataset}
    \label{fig:fashion_dot_dist}
\end{figure}

Figures \ref{fig:cifar10_dot_dist} and \ref{fig:fashion_dot_dist} illustrate the distribution of dot product results on the CIFAR-10 and Fashion-MNIST datasets, respectively. The dot product between the model trained on forgotten data and the server's unlearned model is consistently smaller than that between the model trained on retained data and the unlearned model. This difference is particularly pronounced in the CIFAR-10 dataset. These figures visually demonstrate the underlying reason why the Youden’s Index threshold and the K-means clustering can effectively discriminate the forgotten class: the similarity between models trained on forgotten data and the target unlearned model is systematically lower, creating a statistically separable gap in the parameter space.

\begin{figure}[htbp]
    \centering
    \subfloat[Single-class]{\includegraphics[width=0.45\linewidth]{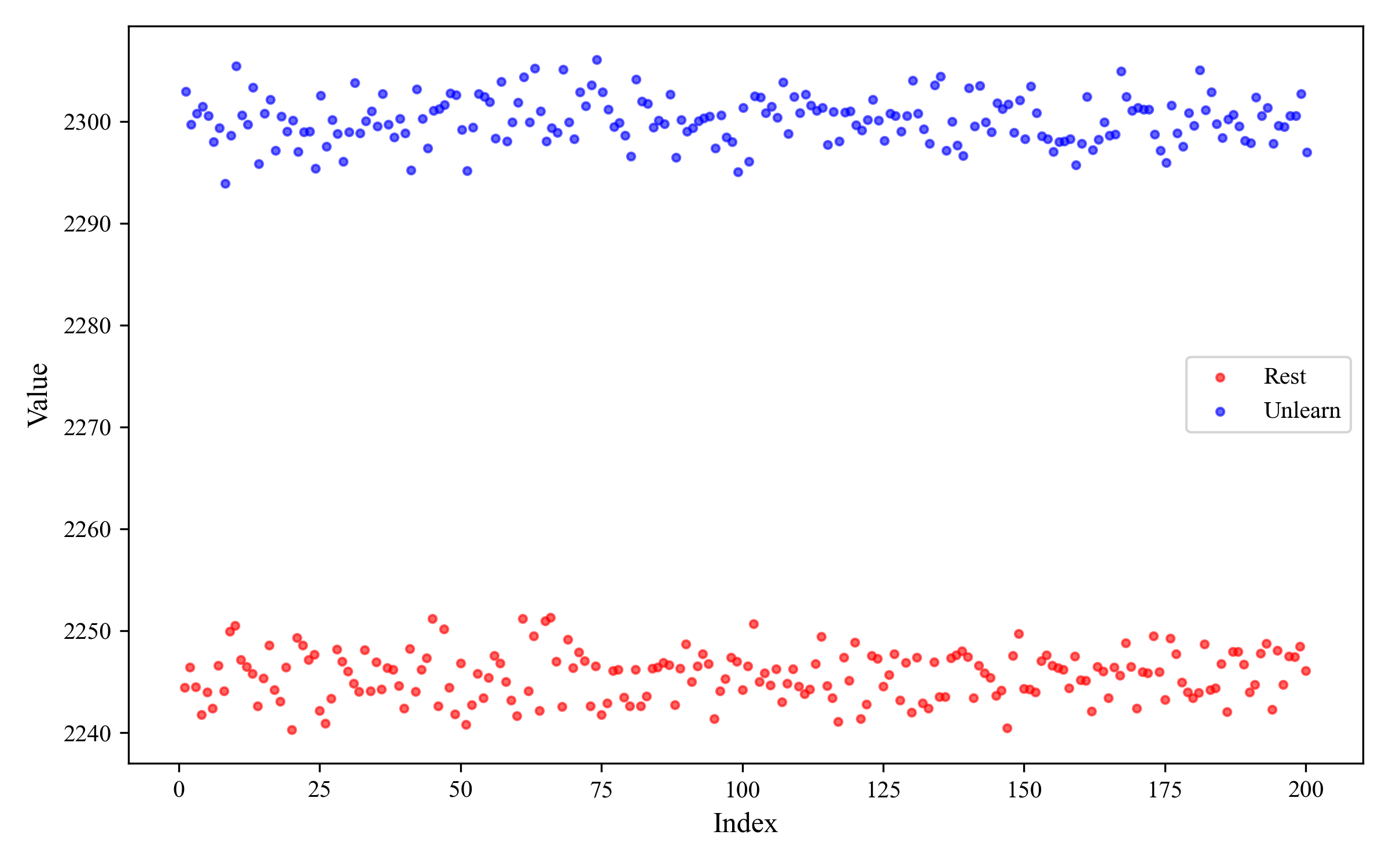}\label{fig:11a}}
    \hfill
    \subfloat[Multi-class]{\includegraphics[width=0.45\linewidth]{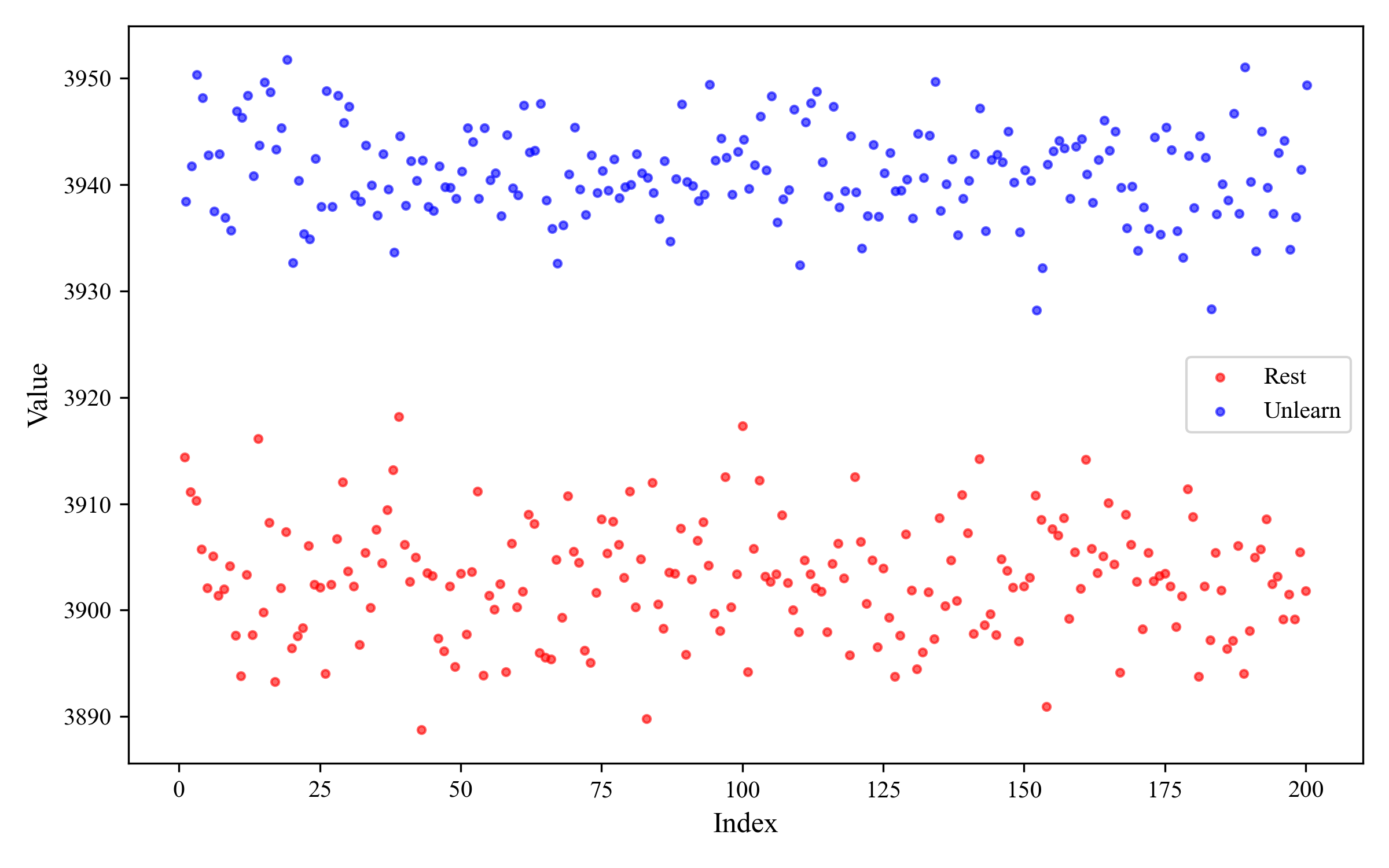}\label{fig:11b}}
    \caption{Distribution of the Summation of Absolute Parameter Differences for Each Model on the SVHN Dataset}
    \label{fig:svhn_param_diff_dist}
\end{figure}

In the white-box attack method based on the Decision Tree, we train multiple models separately on the forgotten and non-forgotten data. The difference vectors between the fully connected layer parameters of these trained models and those of the server model unlearned via retraining constitute the feature set. These features are used to construct a new dataset for training the classifier. To empirically validate the theoretical inequality $\| \mathbf{w}_{\text{rest}} - \mathbf{w}_{\text{unlearn}}' \| > \| \mathbf{w}_{\text{rest}} - \mathbf{w}_{\text{rest}}' \|$, Figure \ref{fig:svhn_param_diff_dist} presents the distribution of the summation of absolute parameter differences. This distribution is calculated for 200 $M_{\text{rest}}'$ models and 200 $M_{\text{unlearn}}'$ models against the server model on the SVHN dataset.

\subsubsection{Unlearning Label Inference via Reversed Labels}
\begin{figure}[htbp]
    \centering
    \subfloat[Single-class]{\includegraphics[width=0.45\linewidth]{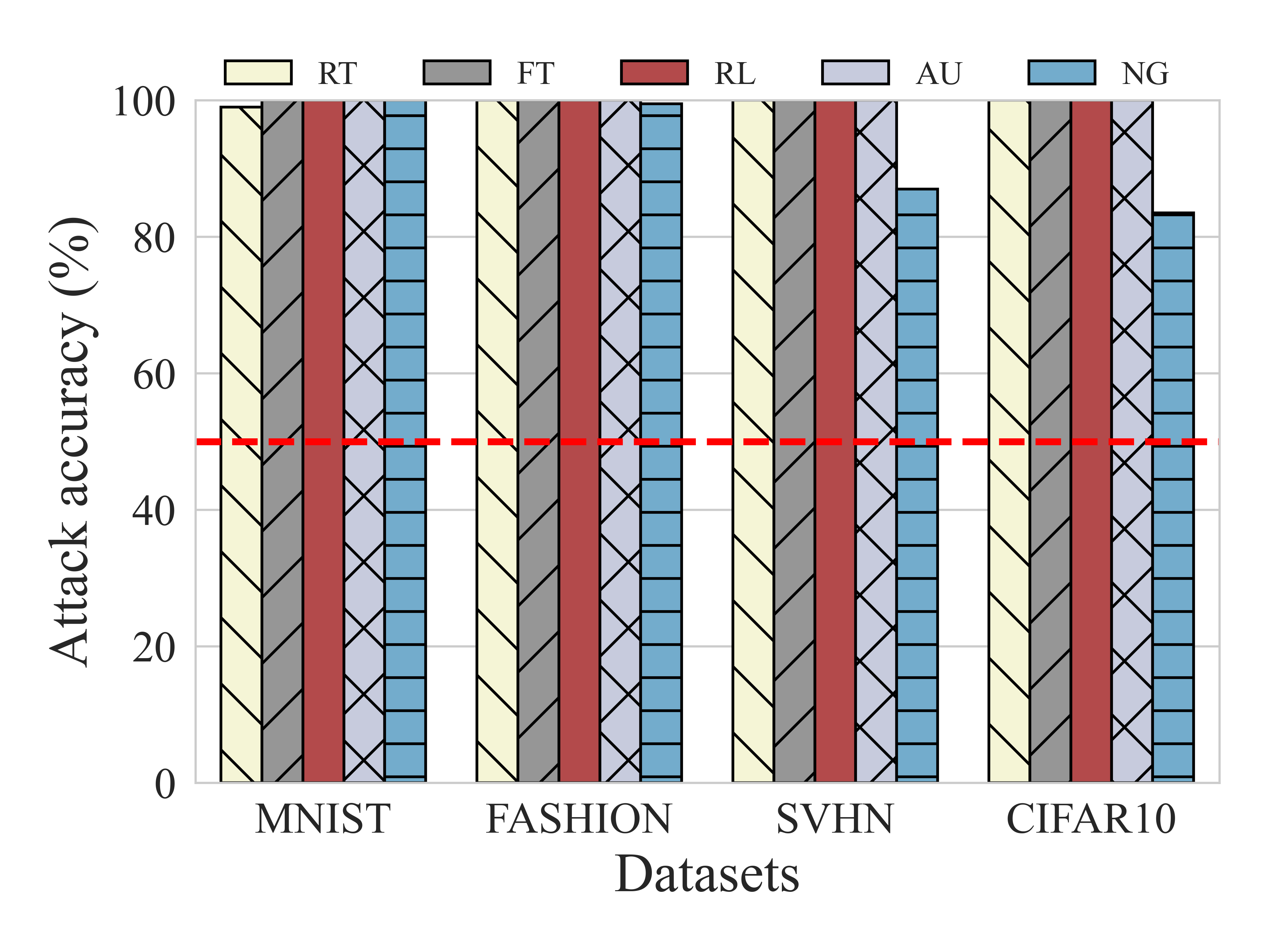}\label{fig:12a}}
    \hfill
    \subfloat[Multi-class]{\includegraphics[width=0.45\linewidth]{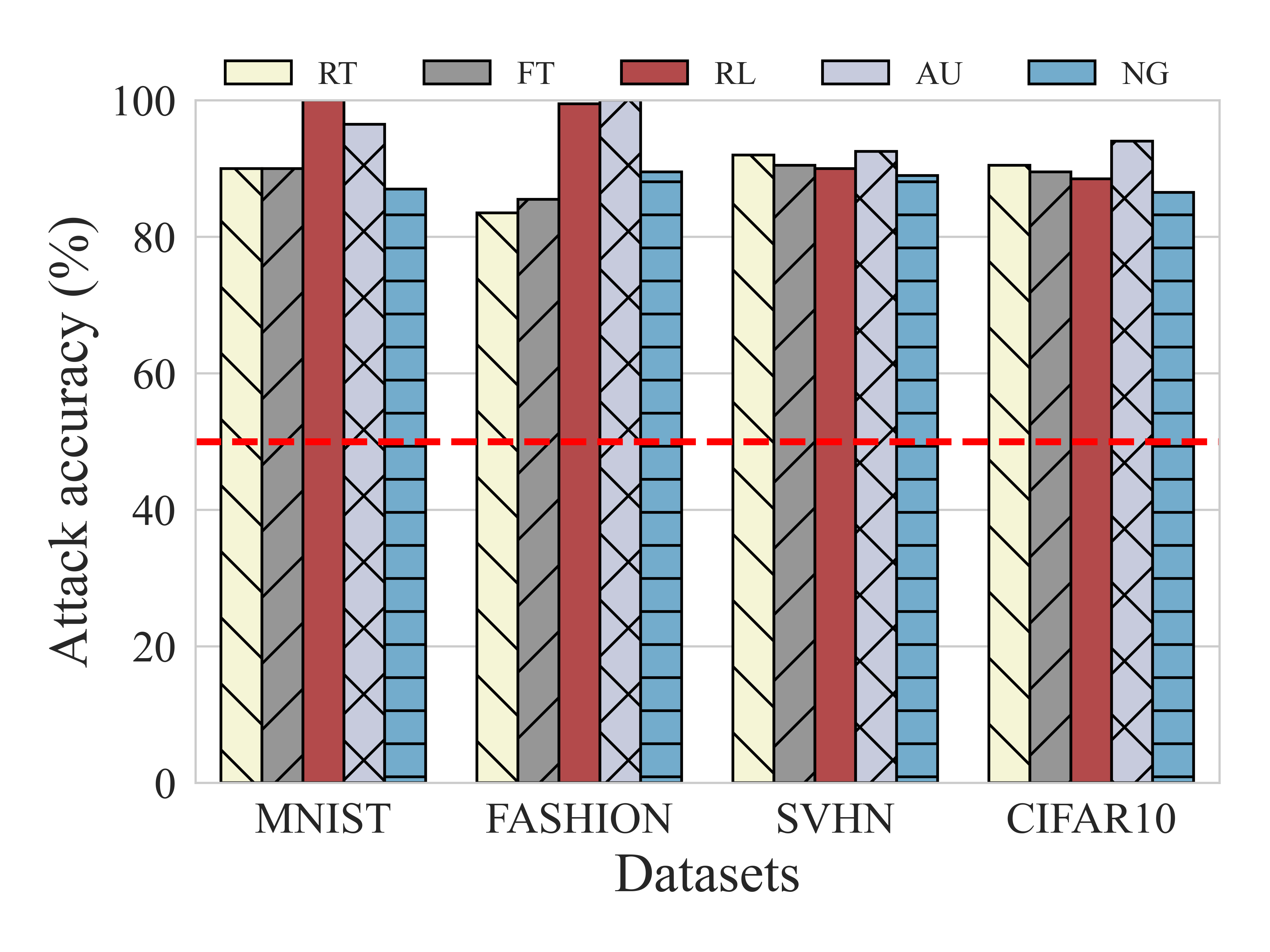}\label{fig:12b}}
    \caption{White-box Attack Results Based on the Threshold Criterion}
    \label{fig:whitebox_threshold_results}
\end{figure}

\begin{figure}[htbp]
    \centering
    \subfloat[Single-class]{\includegraphics[width=0.45\linewidth]{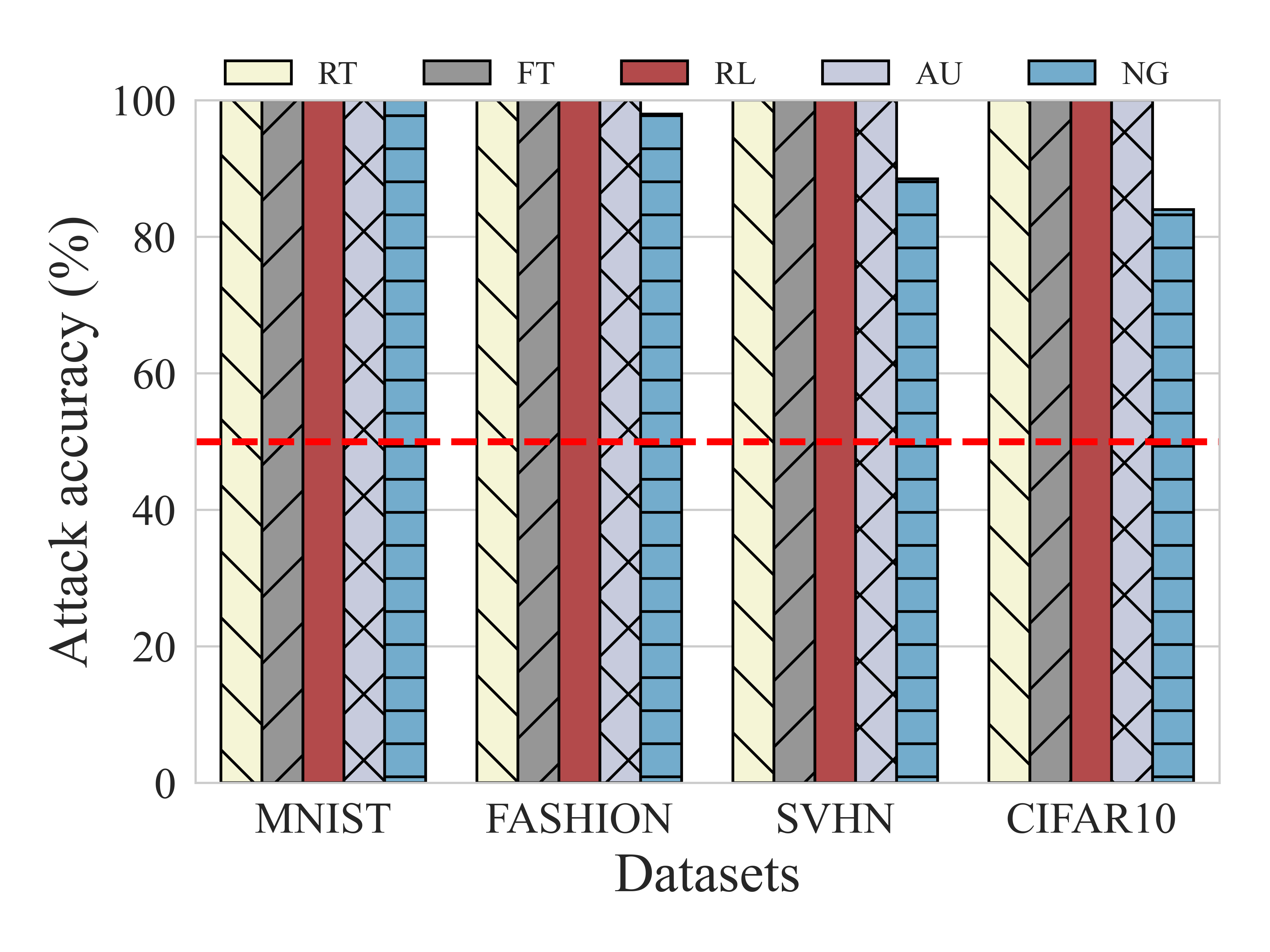}\label{fig:13a}}
    \hfill
    \subfloat[Multi-class]{\includegraphics[width=0.45\linewidth]{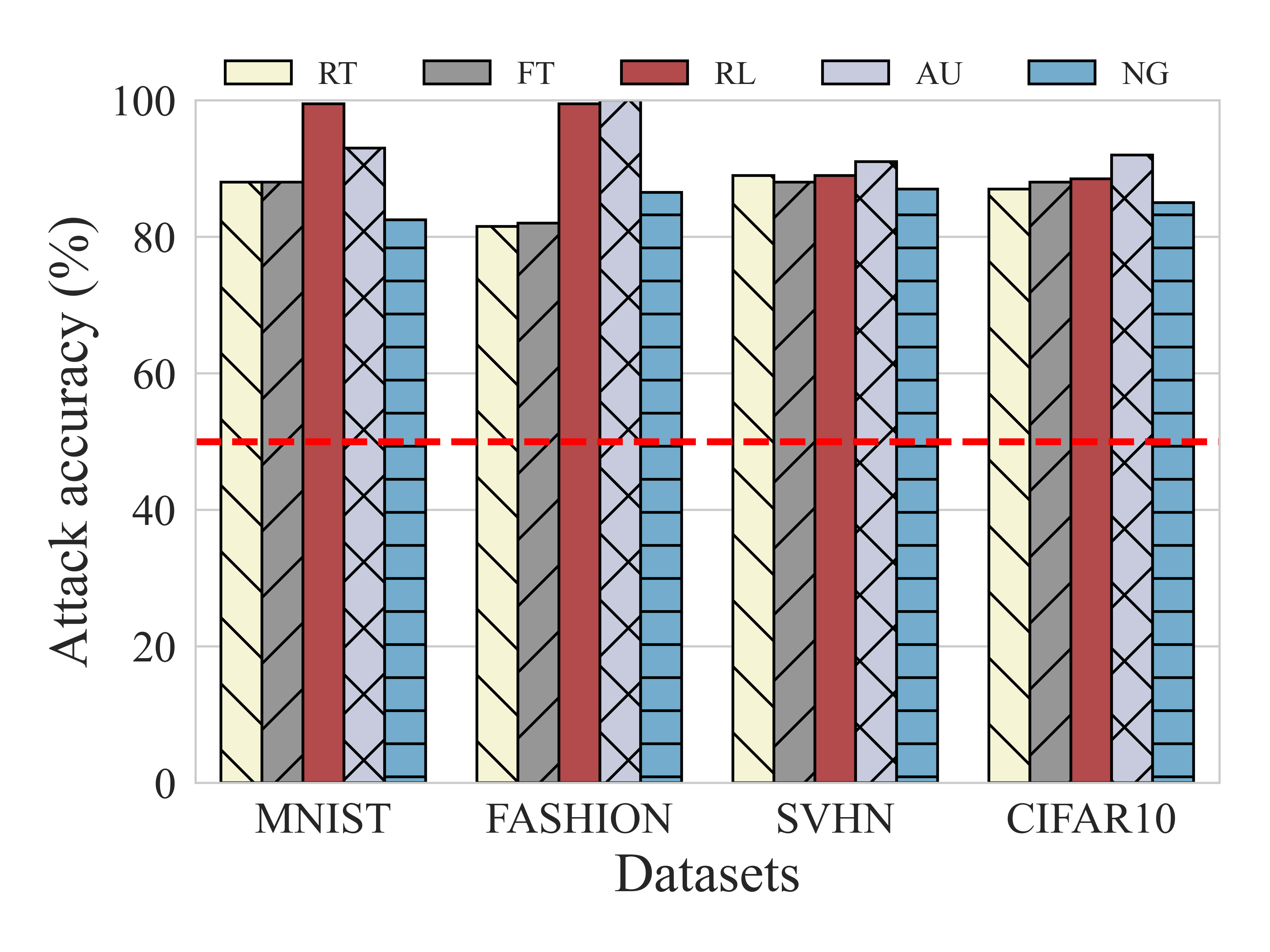}\label{fig:13b}} 
    \caption{White-box Attack Results Based on the Entropy Criterion}
    \label{fig:whitebox_entropy_results}
\end{figure}

Figures \ref{fig:whitebox_threshold_results} and \ref{fig:whitebox_entropy_results} present the white-box attack results based on the threshold criterion and the entropy criterion, respectively, for both single-class and multi-class unlearning scenarios. In terms of classification methodology, the attack performance of the entropy-based criterion is roughly comparable to that of the threshold-based criterion. Regarding the unlearning scenario, the attack effectiveness in the single-class scenario surpasses that in the multi-class scenario. Notably, the proposed method achieves an ASR exceeding 80\% across all datasets.

In both single-class and multi-class unlearning scenarios, the attack success rates of our proposed methods based on both the threshold criterion and the entropy criterion, significantly outperform the baseline method indicated by the red dashed line in the figures. This demonstrates that the methods proposed in this paper can more effectively infer the class labels of forgotten data from the unlearned model.

\subsection{Black-box Attack}
\begin{figure}[htbp]
    \centering
    \subfloat[Single-class]{\includegraphics[width=0.45\linewidth]{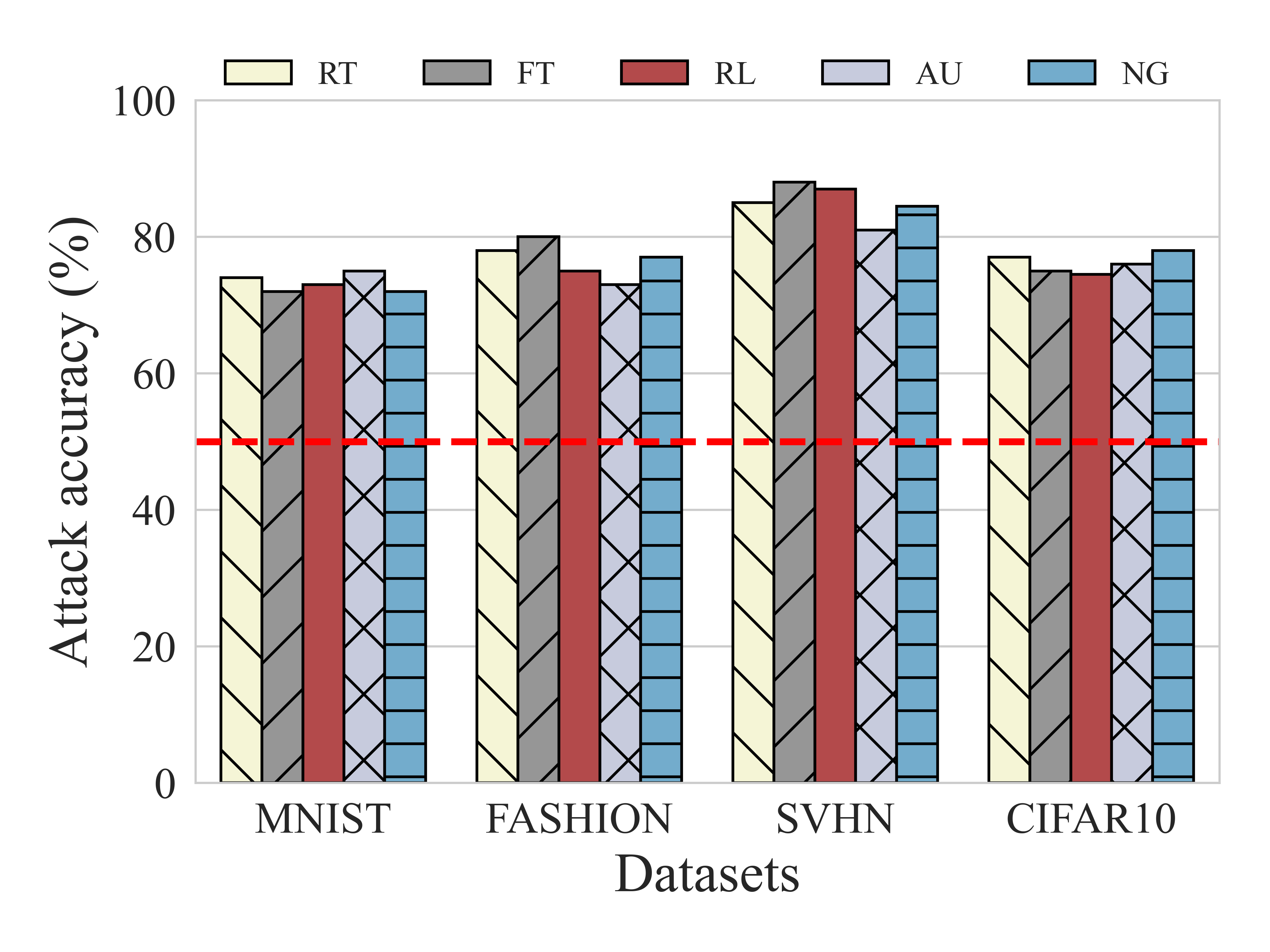}\label{fig:14a}}
    \hfill
    \subfloat[Multi-class]{\includegraphics[width=0.45\linewidth]{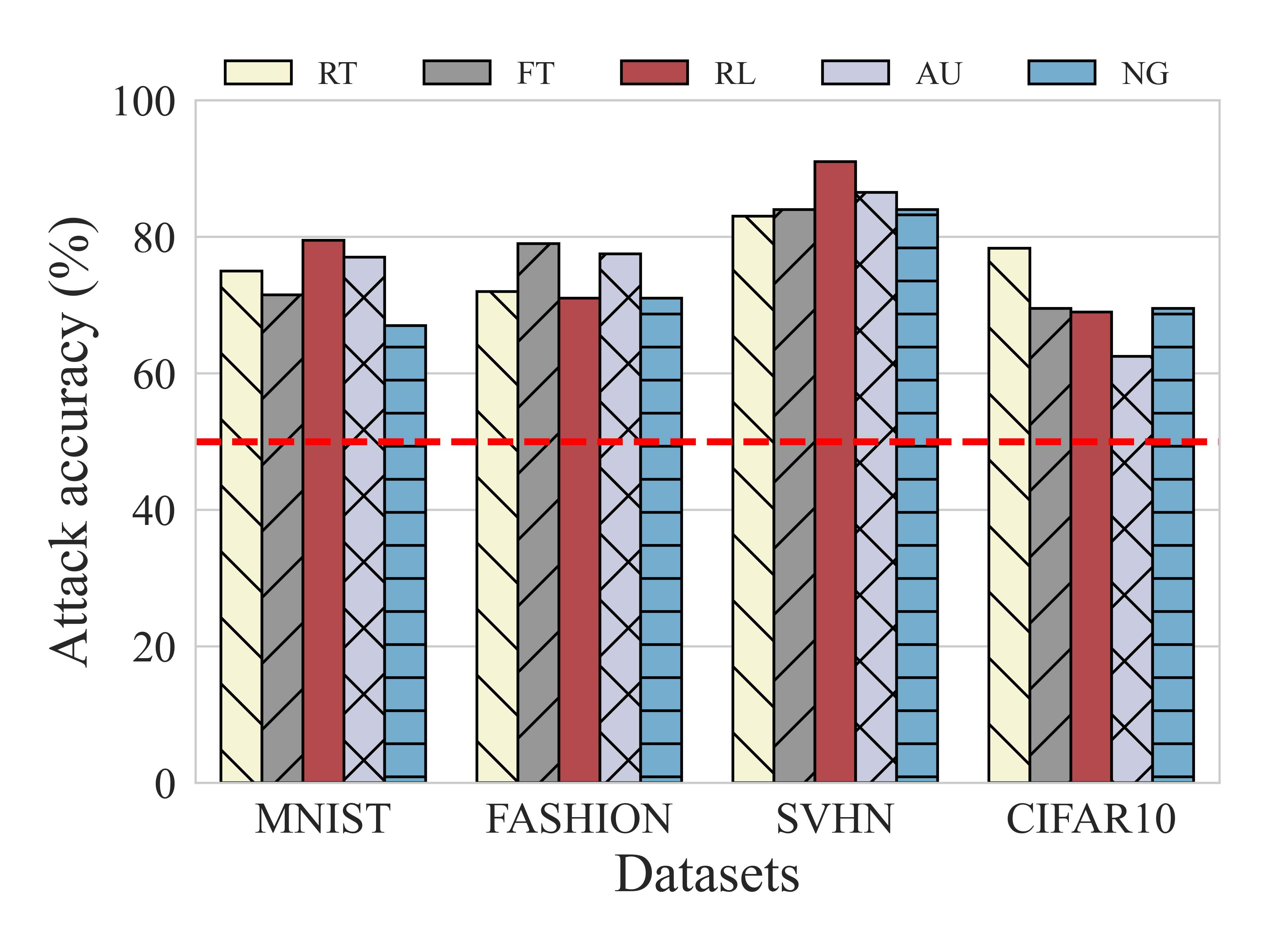}\label{fig:14b}}
    \caption{Black-box Attack Results Based on the Threshold Criterion}
    \label{fig:blackbox_threshold_results}
\end{figure}

\begin{figure}[htbp]
    \centering
    \subfloat[Single-class]{\includegraphics[width=0.45\linewidth]{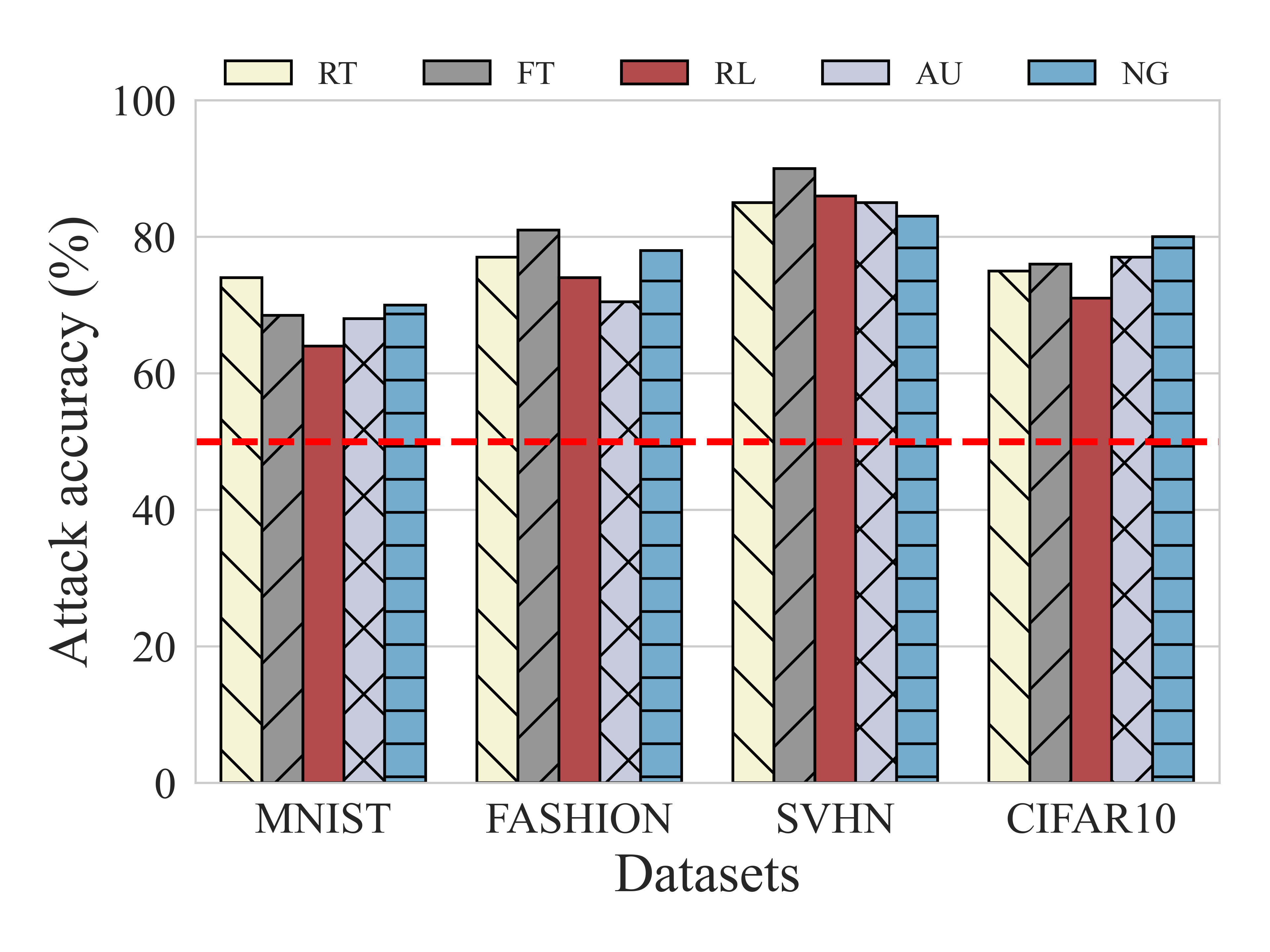}\label{fig:15a}}
    \hfill
    \subfloat[Multi-class]{\includegraphics[width=0.45\linewidth]{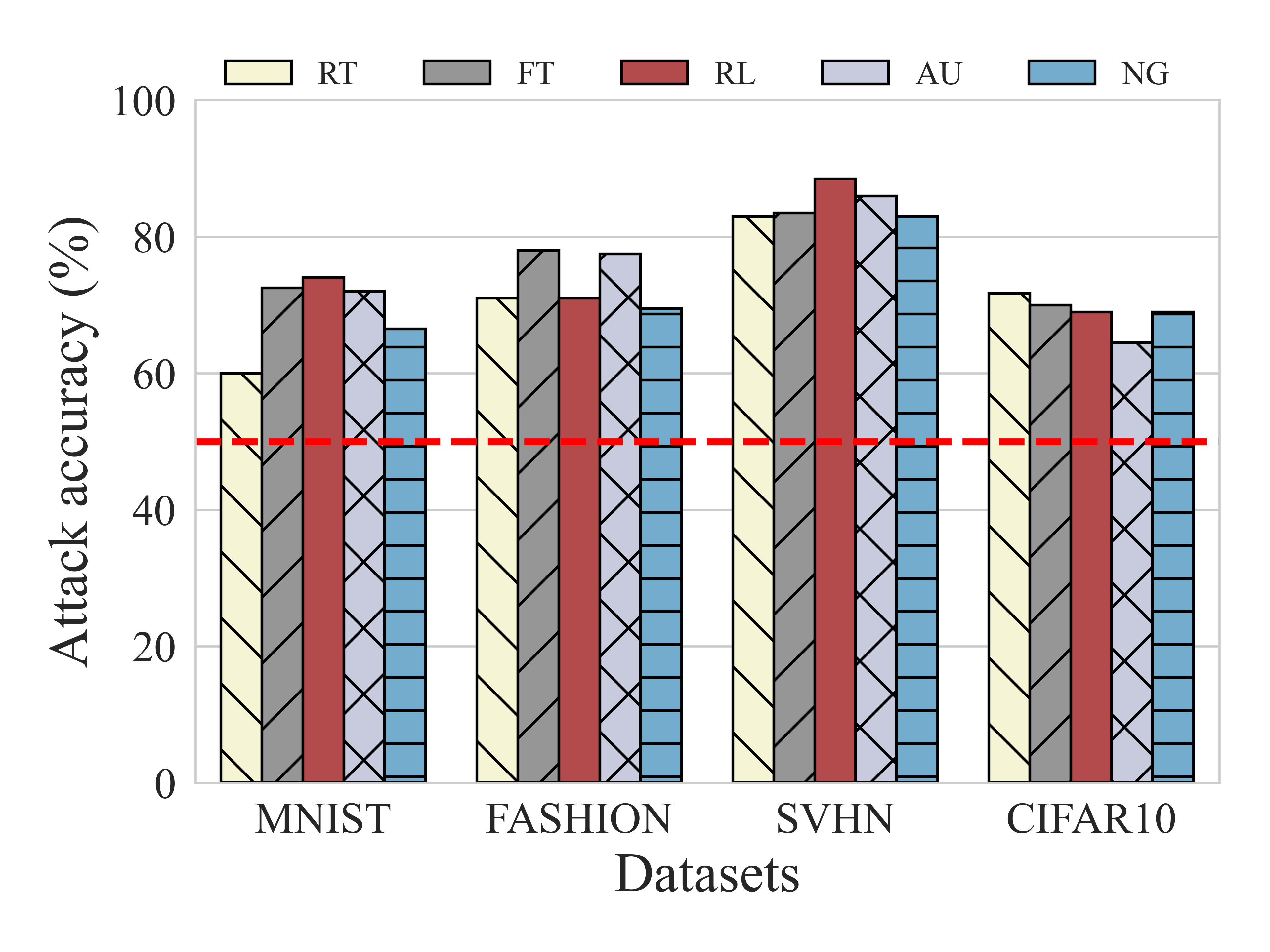}\label{fig:15b}}
    \caption{Black-box Attack Results Based on the Entropy Criterion}
    \label{fig:blackbox_entropy_results}
\end{figure}

Figures \ref{fig:blackbox_threshold_results} and \ref{fig:blackbox_entropy_results} present the black-box attack results based on the threshold criterion and the entropy criterion in single-class and multi-class unlearning scenarios, respectively. The experimental results indicate that although the success rates of black-box attacks are somewhat lower than those of their white-box counterparts, they remain significantly above the random guessing baseline, thereby validating the effectiveness of the proposed attack methods. From the perspective of datasets, the method achieves the best performance on the SVHN dataset. Regarding the scale of unlearned classes, the attack success rate shows no significant difference when the target model forgets one class versus three classes. However, this does not imply that the attack performance remains stable across arbitrary unlearning scales; a more detailed analysis and discussion on this aspect will be provided in the subsequent subsection.

\subsection{Statistical Comparison}
To validate the generalizability of our methods, we conducted a statistical comparison of the average ASR achieved by each attack method across all datasets under different unlearning scenarios.

\begin{table}[htbp]
\centering
\footnotesize
\caption{Statistical Results of Various Attack Methods Based on Model Parameters in the Single-Class Unlearning Scenario}
\label{tab:single_class_attack_results}
\begin{tabular}{@{} c@{\hspace{1pt}} c @{\hspace{1pt}}*{5}{c} @{} c @{}}
\Xhline{1pt}
 \multirow{2}{*}{\makecell{Attack\\Method}} & \multirow{2}{*}{\makecell{Classification\\Method}} &\multicolumn{5}{c}{Average ASR Across All Datasets} & \multirow{2}{*}{\makecell{Overall\\Average}} \\
\cline{3-7}
 & & NG & FT & RL & AU & RT& \\
\hline
\multirow{2}{*}{\makecell{Dot\\Product}} & \makecell{Youden’s\\Threshold} & 81.50 & 84.81 & 89.06 & 84.75 & 90.75 & 86.17 \\
 & K-Means & 80.12 & 83.50 & 88.31 & 83.93 & 90.25 & 85.22\\
\makecell{Difference} & \makecell{Decision Tree} & 90.31 & 90.31 & 92.50 & 90.93 & 93.75  & \textbf{91.56} \\
\Xhline{1pt}
\end{tabular}
\end{table}

\begin{table}[htbp]
\centering
\footnotesize
\caption{Statistical Results of Various Attack Methods Based on Model Parameters in the Multi-Class Unlearning Scenario}
\label{tab:multi_class_attack_results}
\begin{tabular}{@{} c@{\hspace{1pt}} c @{\hspace{1pt}}*{5}{c} @{} c @{}}
\Xhline{1pt}
\multirow{2}{*}{\makecell{Attack\\Method}} & \multirow{2}{*}{\makecell{Classification\\Method}} & \multicolumn{5}{c}{Average ASR Across All Datasets} & \multirow{2}{*}{\makecell{Overall\\Average}} \\
\cline{3-7}
 & & NG & FT & RL & AU & RT \\
\hline
\multirow{2}{*}{\makecell{Dot\\Product}} & \makecell{Youden’s\\Threshold} & 82.62 & 87.0 & 94.0 & 85.93 & 91.25  & 88.16\\
 & K-Means & 81.81 & 85.81 & 93.31 & 85.06 & 90.43  & 87.28\\
\makecell{Difference} & \makecell{Decision Tree} & 95.93 & 96.25 & 96.56 & 96.25 & 95.62 & \textbf{96.12} \\
\Xhline{1pt}
\end{tabular}
\end{table}

Tables \ref{tab:single_class_attack_results} and \ref{tab:multi_class_attack_results} present the average ASR of various attack methods based on model parameters under different unlearning scenarios. The attack method utilizing the dot product of model parameters demonstrates superior performance against unlearning algorithms such as RL and RT, while exhibiting the poorest performance against NG. In contrast, the attack method employing parameter differences consistently delivers strong performance across all unlearning methods, maintaining an ASR above 90\%. This indicates that the parameter difference-based attack method is more effective and robust compared to the dot product-based approach.

\begin{table}[htbp]
\centering
\footnotesize
\caption{Statistical Results of Various Attack Methods Based on Reverse Data in the Single-Class Unlearning Scenario}
\label{tab:single_class_reverse_attack}
\begin{tabular}{@{} c@{\hspace{1pt}} c @{\hspace{1pt}}*{5}{c} @{} c @{}}
\Xhline{1pt}
\multirow{2}{*}{\makecell{Attack\\Method}} & \multirow{2}{*}{\makecell{Classification\\Method}} & \multicolumn{5}{c}{Average ASR Across All Datasets} & \multirow{2}{*}{\makecell{Overall\\Average}} \\
\cline{3-7}
 & & NG & FT & RL & AU & RT & \\
\hline
\multirow{2}{*}{\makecell{White-box\\Attack}} & Threshold & 92.50 & 100.0 & 100.0 & 100.0 & 99.75  & 98.45 \\
 & Entropy & 92.62 & 100.0 & 100.0 & 100.0 & 100.0 & \textbf{98.52}  \\
\multirow{2}{*}{\makecell{Black-box\\Attack}} & Threshold & 77.87 & 78.75 & 77.37 & 76.25 & 78.50 & \textbf{77.75} \\
 & Entropy & 77.75 & 78.87 & 73.75 & 75.13 & 71.42 &76.65 \\
\Xhline{1pt}
\end{tabular}
\end{table}

\begin{table}[htbp]
\centering
\footnotesize
\caption{Statistical Results of Various Attack Methods Based on Reverse Data in the Multi-Class Unlearning Scenario}
\label{tab:multi_class_reverse_attack}
\begin{tabular}{@{} c@{\hspace{1pt}} c @{\hspace{1pt}}*{5}{c} @{} c @{}}
\Xhline{1pt}
\multirow{2}{*}{\makecell{Attack\\Method}} & \multirow{2}{*}{\makecell{Classification\\Method}} & \multicolumn{5}{c}{Average ASR Across All Datasets} & \multirow{2}{*}{\makecell{Overall\\Average}} \\
\cline{3-7}
 & & NG & FT & RL & AU & RT & \\
\hline
\multirow{2}{*}{\makecell{White-box\\Attack}} & Threshold& 88.0 & 88.87 & 94.50 & 95.75 & 89.00  & \textbf{91.22} \\
 & Entropy & 85.25 & 86.50 & 94.12 & 94.00 & 86.37  & 89.24\\
\multirow{2}{*}{\makecell{Black-box\\Attack}} & Threshold & 72.87 & 76.00 & 77.62 & 75.88 & 77.08  & \textbf{75.89} \\
 & Entropy & 72.00 & 76.00 & 75.63 & 75.00 & 71.42  & 74.01 \\
\Xhline{1pt}
\end{tabular}
\end{table}

Tables \ref{tab:single_class_reverse_attack} and \ref{tab:multi_class_reverse_attack} present the average ASR based on inverted data for the same unlearning method across all datasets in the single-class and multi-class unlearning scenarios, respectively. The tables also show the average success rates of attack methods employing different criteria against all unlearning methods. In both single-class and multi-class scenarios, the inversion-based attack methods achieve an average success rate exceeding 70\% in both white-box and black-box settings. The classification method based on the threshold criterion generally slightly outperforms the entropy-based criterion in most cases. Furthermore, compared to the white-box attack based on model parameters and decision trees, the attack method using inverted data with the entropy criterion demonstrates superior performance in the single-class scenario, while showing relatively lower success rates in the multi-class scenario.

\subsection{The impact of the number of forgetting categories}
\begin{figure}[htbp]
\centering
\includegraphics[width=\linewidth]{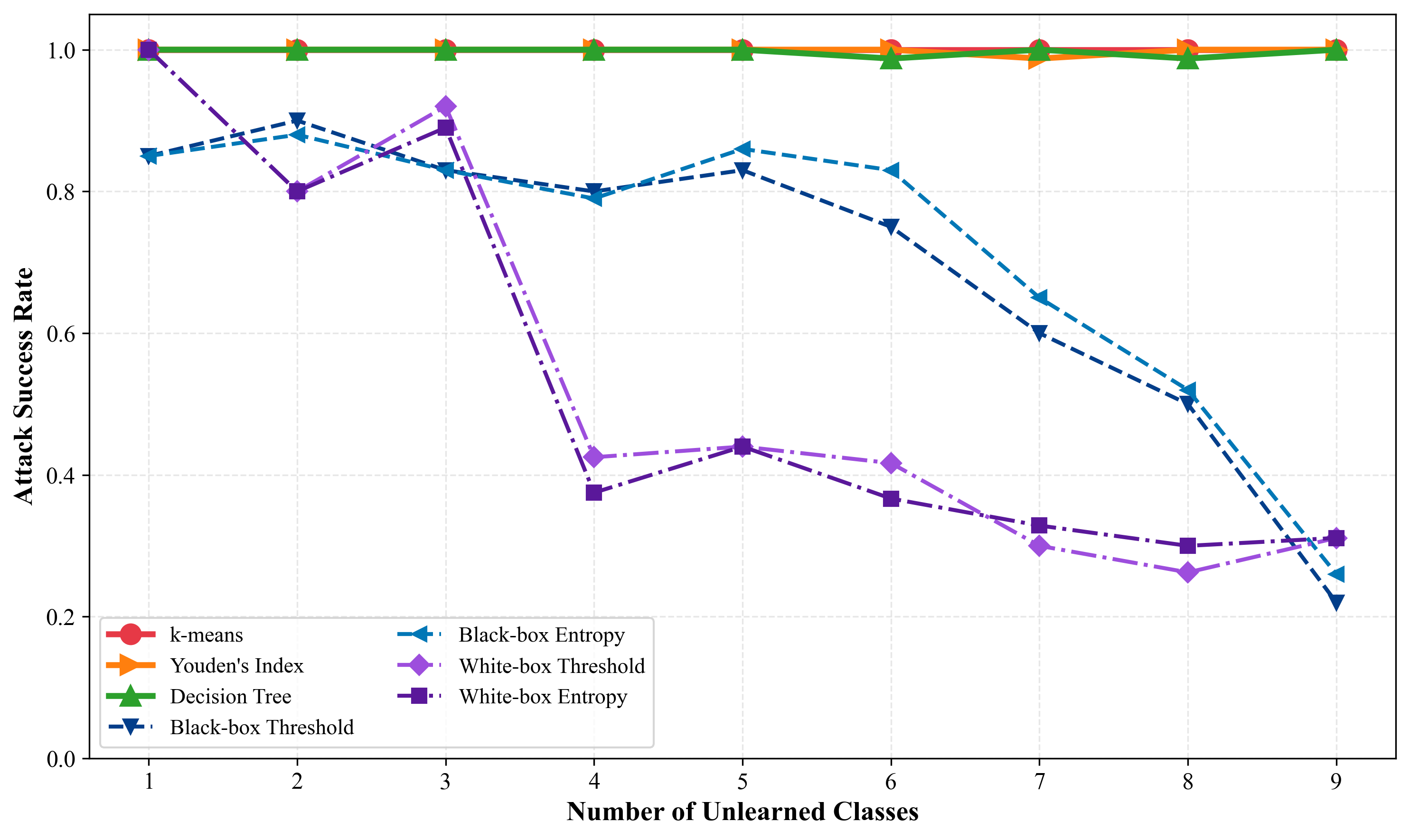} 
\caption{Attack Results on the SVHN Dataset under the Re-Training Unlearning Method with Varying Numbers of Unlearned Classes}
\label{fig:svhn_retrain_unlearn}
\end{figure}

The previous multi-class experiments were conducted with the number of unlearned classes set to 3. To evaluate the robustness of our methods, we performed comprehensive experiments with incrementally increasing numbers of unlearned classes. As shown in Figure \ref{fig:svhn_retrain_unlearn}, the experimental results indicate that the attack method based on data inversion gradually loses effectiveness as the number of unlearned classes increases. Specifically, for white-box attacks, the ASR drops precipitously when the number of unlearned classes exceeds 4. For black-box attacks, a significant decline in ASR begins when the number surpasses 7. In contrast, the attack method based on model parameters consistently maintains a high ASR across different numbers of unlearned classes, demonstrating stable and robust performance overall. This fully illustrates the stability and reliability of the model parameter-based method in complex unlearning scenarios.
\section{Conclusions and discussion}

This paper systematically investigates the critical privacy leakage risks inherent in machine unlearning mechanisms. It reveals vulnerabilities in existing techniques designed to protect data privacy during the forgetting process. The study assumes two realistic attacker profiles: one with access to a limited training dataset and another with no access to any training data. From a theoretical perspective, the paper develops two efficient privacy attack strategies. The first is a white-box attack paradigm based on analyzing model parameters, and the second comprises both white-box and black-box attack schemes based on data inversion. Furthermore, the research integrates screening mechanisms including Youden’s Index, a threshold criterion, and an entropy criterion to achieve efficient identification and filtering of labels associated with forgotten data.

Extensive experiments conducted on four standard datasets and five mainstream unlearning algorithms demonstrate that the proposed methods significantly outperform baseline approaches in terms of ASR under both white-box and black-box settings. Experimental analysis also indicates that, compared to attacks based on data inversion, the model parameter-based attacks exhibit stronger robustness and stability across varying scales of forgotten classes. These findings provide valuable insights for enhancing the security and reliability of model unlearning techniques and for safeguarding individual privacy. The proposed attack methodologies also offer novel perspectives for other privacy attacks in related fields.


\end{document}